\documentclass[12pt]{iopart}

\usepackage{iopams}
\usepackage{xcolor}
\usepackage{subcaption}
\usepackage{braket}
\usepackage{comment}
\usepackage{footmisc}
\expandafter\let\csname equation*\endcsname\relax
\expandafter\let\csname endequation*\endcsname\relax

\usepackage{mathtools} 

\usepackage[normalem]{ulem} 

\makeatletter
\@namedef{ver@amsmath.sty}{}
\makeatother
\usepackage{amstext}

\usepackage[style=phys,%
articletitle=true,biblabel=brackets,%
chaptertitle=false,pageranges=false%
]{biblatex}  
\begin{document}

\title[Artificial atoms from cold bosons in one dimension]{Artificial atoms from cold bosons in one dimension}
\author[Fabian Brauneis, Timothy G Backert, Simeon I Mistakidis, Mikhail Lemeshko, Hans-Werner Hammer, Artem G Volosniev]{Fabian Brauneis$^\dag$, Timothy G Backert$^\dag$, Simeon I Mistakidis$^{\ddag\S}$, 
\\
Mikhail Lemeshko$^\P$, Hans-Werner Hammer$^{\dag\|}$, and Artem G Volosniev$^\P$}

\address{\dag~Technische Universit\"{a}t Darmstadt, Department of Physics, 64289 Darmstadt, Germany}
\address{\ddag~ITAMP,  Center for Astrophysics $|$ Harvard $\&$ Smithsonian, Cambridge, MA 02138 USA}
\address{\S~Department of Physics, Harvard University, Cambridge, Massachusetts 02138, USA}
\address{$\|$~ExtreMe Matter Institute EMMI and Helmholtz Forschungsakademie
  Hessen f\"ur FAIR (HFHF), GSI Helmholtzzentrum f\"ur Schwerionenforschung GmbH, 64291 Darmstadt, Germany}
  \address{\P~Institute of Science and Technology Austria, Am Campus 1, 3400 Klosterneuburg, Austria}
\date{September 2021}

\begin{abstract}
We investigate the ground-state properties of weakly repulsive
one-dimensional
bosons in the presence of an attractive zero-range impurity potential.
First,
we derive mean-field solutions to the problem on a finite ring for the two asymptotic cases:
(i) all bosons are bound to the impurity and (ii) all bosons are in a
scattering state. Moreover, we derive the critical line that separates
these
regimes in the parameter space. In the thermodynamic limit, this
critical line
determines the maximum number of bosons that can be bound by the
impurity potential, forming an artificial atom. Second, we validate the
mean-field results using the flow equation approach and the multi-layer
multi-configuration time-dependent Hartree method for atomic mixtures.
While
beyond-mean-field effects destroy long-range order in the Bose gas, the
critical boson number is unaffected. Our findings are important for
understanding such artificial atoms in low-density Bose gases with
static and
mobile impurities.
\end{abstract}

\maketitle

\newpage

\section{Introduction}

A quantum state is bound if the probability to
 find parts of the system infinitely far from each other vanishes. 
 It is one of the basic problems in quantum mechanics to determine conditions for a bound state to occur. Such problems are typically encountered in few-body settings. However, they also play an important role in many-body physics. For example, a low-energy model of dilute many-body systems may include  bound states as building blocks.
 
 Only in some special cases, there exist results that provide conditions for binding. For example, any attractive potential supports a two-body bound state in one (1D) and two (2D) spatial dimensions, whereas solely `deep' potentials can lead to a bound state in three dimensions (3D)~\cite{landau1977quantum, Simon1976}. For more than two particles, general conditions for binding are not known. Moreover, there seem to be no universal theoretical approaches to find them. Typically, one has to resort to numerical calculations, and only some problems can be addressed analytically (within certain approximation schemes). The latter class of problems includes for example the Efimov effect, which provides a universal mechanism for resonant interactions in 3D~\cite{Efimov:1970zz,Nielsen2001, Braaten:2004rn,Naidon:2016dpf}. Another example of analytically tractable model are bound states of weakly repulsive bosons attracted by a short-range potential~\cite{Kolomeisky2004,Straley2007,TrappingCollapse,QuantumBlockade}. That system is reminiscent of an atom where the role of electrons is played by the bosons, and the nucleus is realized by the potential. Therefore, in what follows we shall occasionally refer to the system as an `artificial atom from bosons'.

Artificial atoms were mainly studied in one or three spatial dimensions. 
In 3D, different theoretical methods seem to disagree on the number of bosons that can be bound to an impurity~\cite{Kolomeisky2004,Straley2007,TrappingCollapse,QuantumBlockade}. In 1D, there is a similar puzzle.  The outcome of the mean-field approximation~\cite{Gunn,Kolomeisky2004} is not supported by the phenomenological argument of Ref.~\cite{TrappingCollapse}. The latter study argues that 
a dilute Bose gas can always be mapped onto a system of non-interacting fermions implying that only a single boson can be bound (cf. the Pauli exclusion principle). However, the mean-field studies demonstrate that the number of bound bosons can be large if boson-boson interactions are weak. These different results certainly motivate further investigations of the `artificial atom' problem. Besides providing insight into conditions for binding, they shed light onto the physics of the Bose polaron (see, e.g.,~\cite{Catani2012,Jorgensen2016,Hu2016}), in particular onto the polaron-to-molecule transition region.
In this paper, we focus on properties of a one-dimensional artificial atom.
\\
\\
{\it Main results of the paper} 
\\
\\ Our first result concerns
\begin{itemize}
    \item {\it the mean-field solution of an artificial-atom problem in a finite ring}. This solution rigorously shows that an impurity can support a many-boson bound state.  
\end{itemize}
Specifically, the mean-field solution unveils the existence of three different physical scenarios depending on the number of bosons, $N$, see Eqs.~(\ref{eq:MFmb}) and~(\ref{eq:MFscatt}).
Below a critical particle number all bosons are trapped by the impurity, which confirms the previous findings of Refs.~\cite{Gunn,Kolomeisky2004}.
 In this regime, the density of the Bose gas decays exponentially, see Eq.~(\ref{eq:mbb_tail}). Therefore, we classify the system as a {\it many-body bound state}. At the critical particle number, all bosons are also bound to the impurity. However, the corresponding density decays as $1/x^2$, see Eq.~(\ref{eq:MfCritLlarge}). Therefore, we shall say that the system is in a {\it critical state}. If the number of bosons is larger than the critical one, a certain portion of bosons occupies scattering states, i.e., there is a significant probability to find a boson far from the impurity.  
   \\
   \\
The second result of this paper is 
\begin{itemize}
    \item {\it a validation of the mean-field predictions for the artificial atom problem using numerical beyond-mean-field methods.}
\end{itemize}
To this end, we use a recently introduced in-medium similarity renormalization group method for bosons (IM-SRG; also called flow equations)~\cite{Volosniev2017} whose accuracy is confirmed here using the well-established multi-layer multi-configuration time-dependent  Hartree  method for atomic mixtures (ML-MCTDHX)~\cite{cao2017unified}. These methods allow us to study the decay of phase correlations and demonstrate phase coherence between the bosons in the vicinity of the impurity, see Fig.~\ref{fig:Ltoinfty:phasefluc}. We conclude that the mean-field solution describes the system well as long as all bosons are bound to the impurity. When bosons populate scattering states, they occupy the whole space, which lowers the density and increases phase fluctuations. The here employed numerical methods can be used to test the argument of Ref.~\cite{TrappingCollapse} that bosons fermionize in artificial atoms in 1D. Our results suggest that fermionization occurs only for bosons in scattering states. 
\\
\\
{\it Structure of the paper}
\\
\\
The paper is structured as follows: Section~\ref{chap:sys} introduces the system under consideration. Section~\ref{chap:mf} presents the mean-field solution, which is further analyzed in the zero-density limit in Sec.~\ref{chap:Expand}. Further, the mean-field solution is benchmarked against the flow equations results in Sec.~\ref{Chap:IMSRGZeroDensity}.
A mobile impurity in a Bose gas is studied in Sec.~\ref{chap:mobile}; it is concluded that the mean-field approach describes that system also well. Section~\ref{sec:concl} contains a brief summary and outlook.  For convenience, we provide five appendices that elaborate on technical details of our study. \ref{App:NumMethods}  describes the employed numerical methods. They are benchmarked against one another in \ref{App:ComparisonSimos}. \ref{App:Closed} presents a mean-field solution for a box trap. \ref{App:DetailsZeroDensity} contains some details on the mean-field solution in the zero-density limit. In \ref{App:Few-body_Limit} we discuss the smallest non-trivial system -- a two-boson artificial atom.

\section{System: An Impurity Atom in a Bose Gas}
\label{chap:sys}

\subsection{Hamiltonian}
We investigate a one-dimensional system of $N$ bosons and an impurity in a ring. The standard Hamiltonian for such a system in the context of cold-atom experiments reads (see, e.g.,~\cite{Sowinski2019,Mistakidis2022} and references therein)
\begin{equation}
    H=-\frac{1}{2m}\frac{\partial ^2}{\partial y^2}-\frac{1}{2M}\sum\limits_{i=1}^N\frac{\partial^2}{\partial x_i^2}+c\sum\limits_{i=1}^N\delta(x_i-y)+g\sum\limits_{i,j}\delta(x_i-x_j)\,,
    \label{eq:PolaronHam}
\end{equation}    
where we assume $\hbar=1$; $y$ ($x_i$) is the position of the impurity ($i$th boson), $m$ refers to the mass of the impurity atom, and $M$ denotes the mass of a boson.  For convenience, we shall use a system of units with $M=1$ in the numerical calculations. To model atom-atom interactions, we employ delta-function potentials that describe $s$-wave scattering~\cite{Olshanii1998}, which is dominant in the ultracold regime. Their strengths $c$ and $g$ can be virtually arbitrary thanks to the possibility to tune them via external fields using the phenomenon of Feshbach resonances~\cite{Chin2010}.
  
\begin{figure}[t]
    \centering
    \centering
    \includegraphics[width=1\linewidth]{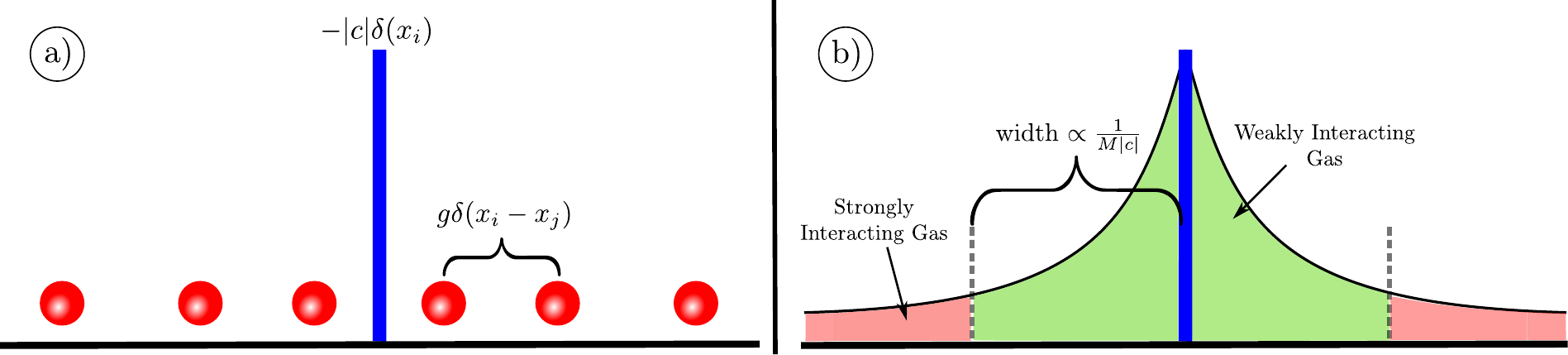}
    \caption{Panel a): Illustration of the system. Red balls represent $N$ bosons. The blue vertical line is the heavy attractive impurity. Panel b): Sketch of the density of the Bose gas for a finite value of $N$, and a large system size (i.e., zero-density limit) assuming that $g/|c|\ll1$. Near the impurity, at distances $\sim 1/(M c)$, the density of the Bose gas is high due to the impurity-boson attraction and thus the effective boson-boson interaction is weak. For larger distances from the impurity, the density is low, which implies that the Bose gas is strongly interacting there. [Note that it is specific to 1D systems that a lower density corresponds to stronger interactions. For example, in three dimensions, the situation is reversed -- low densities imply weak interactions.]}
    \label{fig:OneImpurity:SketchSystem}
\end{figure}

For simplicity, we first focus on a heavy impurity, $m/M\rightarrow\infty$; the role of the impurity mass is briefly discussed in Sec.~\ref{chap:mobile}. Without loss of generality, we place the impurity at $y=0$ as illustrated in Fig.~\ref{fig:OneImpurity:SketchSystem}~a). Note that from the experimental point of view, a heavy impurity can be realized using atoms with very different masses, e.g., $^{7}$Li (bosons) and $^{174}$Yb (impurity)~\cite{Schafer2018}, such that the kinetic energy of the impurity can be neglected. Alternatively, a localized external field (light blade) can be used to trap the impurity atom (see, e.g.,~Ref.~\cite{Catani2012}) or to even simply produce a delta-function potential, see, e.g., Ref.~\cite{Lebrat2019}. Note that these different experimental methods might lead to different finite range effects whose investigation we leave for future studies.

Below, we focus on attractive boson-impurity ($c<0$) and repulsive boson-boson interactions ($g>0$). In the main part of the discussion, we consider periodic boundary conditions, i.e., particles are confined to a ring of length $L$. For completeness, we also present results for a box trap in \ref{App:ComparisonSimos} and \ref{App:Closed}. Note that such boundary conditions can also be realized in experiment, see, e.g., Refs.~\cite{Bell_2016, Mukherjee2017}.

\subsection{Physical Picture}
\label{subsec:phys_int}

Before we proceed with an analysis of the Hamiltonian $H$, let us provide some basic insight into the physics of the system, which is driven by the interplay between attractive impurity-boson and repulsive boson-boson interactions. 
First we note that a delta-function potential supports a single bound state (see, e.g.,~\cite{Griffiths2005}). This means that an arbitrary number of bosons can be trapped by the impurity if the bosons are non-interacting ($g=0$). In contrast, if $1/g=0$, the bosons fermionize~\cite{Girardeau1960}, and only one boson can be trapped by the impurity. [Indeed, only one fermion can be trapped by the impurity due to the Pauli exclusion principle]. This observation implies that the interplay between the attractive impurity-boson interaction and repulsive boson-boson interaction should lead to a critical number of bosons, $N_{\mathrm{cr}}$, that can be bound to an impurity. 

In this work, we estimate this critical number $N_{\mathrm{cr}}$ from a mean-field approximation (see also Refs.~\cite{Gunn, Kolomeisky2004}). We also investigate the system using two numerical approaches, namely the IM-SRG~\cite{Volosniev2017} and the ML-MCTDHX~\cite{cao2017unified} (for a brief description of these methods see \ref{App:NumMethods}). These methods allow us to estimate the importance of beyond-mean-field effects from the decay of the quasi-long-range order as captured by the system's reduced density matrix, see Sec. \ref{Chap:IMSRGZeroDensity} and~\ref{App:ComparisonSimos}. 

To understand why the mean-field approximation is accurate, let us consider the system in the limit $L\to\infty$ [$N$ is fixed] and $g/|c|\ll1$, which is one of the main limits of this work. 
The effective strength of the boson-boson interactions can be parameterized by $M g/\rho(x)$, where $\rho(x)$ is the density of the Bose gas.  This parameterization is natural for 1D problems, see, e.g.,~\cite{Pethick2002}. The value of $M g/\rho(x)$ is the smallest in the vicinity of the impurity, and it grows towards the edge of the bound state. For example for $g=0$, $\rho(x)= N |c|M e^{-2M|cx|}$, see, e.g., Ref.~\cite{Griffiths2005}. Assuming that this density approximates also the system with $g/|c|\ll 1$, we conclude that 
\begin{equation}
    \frac{g}{\rho(x)}\simeq \frac{g}{N M|c|}e^{2M |cx|}. 
    \label{eq:effective_physical}
\end{equation}
 Therefore, the mean-field ansatz must describe the Bose gas well in the vicinity of the impurity as long as $x$ is not large.  
The characteristic width of this `mean-field' region is proportional to $1/(M |c|)$. Farther away from the impurity, the density of the Bose gas is low, hence, the boson-boson interactions are strong, and the mean-field ansatz is no longer applicable, see Fig.~\ref{fig:OneImpurity:SketchSystem}~b). We extend this line of argumentation in Sec.~\ref{chap:Expand}.

\subsection{Relevant Length Scales}

Three parameters define the length scales in our model: $1/(Mg)$, $1/(M |c|)$ and $L$. 
One can always employ one of them to define the system of units. In our work, it is convenient to use $1/(M|c|)$ for this purpose because it is the only relevant length scale for the impurity-boson bound state if $N=1$ or $g=0$ (see Eq.~(\ref{eq:effective_physical})). The corresponding two dimensionless parameters are: the relative interaction strength $\alpha=c/g$ and the dimensionless length $LM|c|$.
Note that the latter will be useful sometimes to express as $LM|c|/N$ or equivalently as $M|c|/\rho$, where $\rho=N/L$ is the density of the Bose gas without the impurity. This will be especially convenient 
in Sec.~\ref{Chap:IMSRGZeroDensity} where we consider how the system approaches the zero-density limit ($L\to\infty$ and $N$ is finite).

\section{Mean-Field Solution for the Heavy Impurity Problem}
\label{chap:mf}

For a system consisting of weakly interacting bosons it is reasonable to assume that the ground-state wave function is a product state: $\Phi=\prod\limits_i f(x_i)$. Here, $f(x)$ is a single-particle function obtained by minimizing $\langle\Phi|H|\Phi\rangle$. This minimization procedure leads to the Gross-Pitaevskii equation (GPE)~\cite{Pethick2002}:
\begin{equation}
    -\frac{1}{2M}\frac{d^2f}{dx^2}+g(N-1)f(x)^3+c\delta(x)f(x)=\mu f(x)\,,
    \label{eq:GPE}
\end{equation}
where $\mu$ is the chemical potential\footnote{It is interesting to note that this equation was also derived and studied for a heavy atom in a strong magnetic field, see parameter regime 5 (`region 5') of Ref.~\cite{Lieb1992}.}. By assumption, the function $f$ is periodic i.e. $f(-L/2)=f(L/2)$. [For a brief discussion of a system in a box trap where $f(-L/2)=f(L/2)=0$, see~\ref{App:Closed}]. Note that some care is needed when using a mean-field approximation in 1D where quantum fluctuations destroy the condensate in the thermodynamic limit~\cite{Kane1967,Popov1972, NoBECHohenberg, Pethick2002}. We shall rely on {\it ab-initio} numerical calculations to confirm that the mean-field approximation is indeed accurate, at least for describing the Bose gas in the vicinity of the impurity. The relevant physical picture is given in Sec.~\ref{subsec:phys_int}.  

We notice that Eq.~(\ref{eq:GPE}) with $c=0$ is integrable, see, e.g., Ref.~\cite{abramowitz1972handbook}. Therefore, one can follow the same strategy as when solving the Schr{\"o}dinger equation with the delta-function interaction~\cite{Griffiths2005}, i.e., use the known solutions for $x>0$ and $x<0$, then implement the boson-impurity interaction $c\delta(x)$ as the boundary condition at $x=0$. 
Once the mean-field solution is obtained, it is possible to calculate any observable of interest. For example, the energy of the system is determined by

\begin{equation*}
    \frac{E}{N}=\mu-
\frac{g(N-1)}{2}\int\limits_{-L/2}^{L/2} |f(x)|^4 \mathrm{d}x.
\end{equation*}

Below, we present the two solutions to Eq.~(\ref{eq:GPE}) that, by increasing $L$ to infinity, connect adiabatically to the two different physical situations: (i) all bosons are bound to the impurity, (ii) no boson is bound, see the next section. The two solutions coincide at the threshold for binding, which we refer to as the point of transition (PoT). Note that the bound-state solution was discussed in Refs.~\cite{Gunn,Kolomeisky2004} for $L\to\infty$. 

We focus on systems with finite values of $L$, since they allow us to directly benchmark the mean-field method against beyond-mean-field numerical approaches. In addition, our solution is relevant to cold-atom experiments, which typically have a finite size. Last but not least, the finite-$L$ solution provides insight into the case with $N>N_{\mathrm{cr}}$, which is important for understanding the thermodynamic limit, as we plan to discuss in an upcoming work.

\subsection{Mean-field Solutions}

The first solution to Eq.~(\ref{eq:GPE}) reads
\begin{equation}
    f_{\mathrm{mbb}}(x)=\sqrt{\frac{4K(p_{\mathrm{mbb}})^2}{M gL^2\delta^2(N-1)}}\text{ds}\left(2K(p_{\mathrm{mbb}})\left[\frac{|x|}{\delta L}+\frac{1}{2}-\frac{1}{2\delta}\right],p_{\mathrm{mbb}}\right).
    \label{eq:MFmb}
\end{equation}
It is determined by  the Jacobi elliptic function ds \cite{abramowitz1972handbook}.
By construction, this solution is parity symmetric $f_{\mathrm{mbb}}(-x)=f_{\mathrm{mbb}}(x)$. 
The chemical potential is 
\begin{equation*}
    \mu_{\mathrm{mbb}}=\frac{2K(p_{\mathrm{mbb}})^2(1-2p_{\mathrm{mbb}})}{M\delta^2L^2},
\end{equation*}
where $K$ is the complete elliptic integral of the first kind.

The second solution to Eq.~(\ref{eq:GPE}) is given by the Jacobi elliptic function ns:
\begin{equation}
    f_{\mathrm{scatt}}(x)=\sqrt{\frac{4K(p_{\mathrm{scatt}})^2}{M gL^2\delta^2(N-1)}}\text{ns}\left(2K(p_{\mathrm{scatt}})\left[\frac{x}{\delta L}+\frac{1}{2}-\frac{1}{2\delta}\right],p_{\mathrm{scatt}}\right).
    \label{eq:MFscatt}
\end{equation}
The corresponding chemical potential is
\begin{equation*}
    \mu_{\mathrm{scatt}}=\frac{2K(p_{\mathrm{scatt}})^2(p_{\mathrm{scatt}}+1)}{M\delta^2L^2}.
\end{equation*}

The parameters $p_{\mathrm{scatt}}\in[0,1)$, $p_{\mathrm{mbb}}\in[0,1)$ and $\delta$\footnote{Note that the solutions $f_{\mathrm{scatt}}$ and $f_{\mathrm{mbb}}$ can be transformed into one another via
    $p_{\mathrm{mbb}}=-\frac{p_{\mathrm{scatt}}}{1-p_{\mathrm{scatt}}}$
if one allows for negative values of $p$. The parameter $\delta$ remains unchanged in this transformation.} that enter in the definitions above are fixed by normalization, and the boundary condition due to the impurity-boson potential
\begin{equation*}
\label{eq:NormCon}
    \int\limits_{-L/2}^{L/2}|f(x)|^2\mathrm{d}x=1, \qquad 
    \frac{df}{dx}\bigg|_{x\to 0+}=M cf(0)\,.
\end{equation*}

It is straightforward to write these conditions in a more explicit form. For example, for $f_{\mathrm{mbb}}$, the normalization condition leads to 
\begin{align}
    \mathcal{E}\left(K-\frac{K}{\delta},p\right)+(1-p)\mathrm{nd}\left(\frac{K}{\delta},p\right)\mathrm{sc}\left(\frac{K}{\delta},p\right)-E(p)+\frac{1-p}{\delta}K=\frac{MgL\delta(N-1)}{4K},
    \label{eq:norm_condition_mbb}
\end{align}
where  $K=K(p)$, and we imply that $p=p_{\mathrm{mbb}}$ [$\mathcal{E}$, $E(p)$ (not to be confused with the energy), $\mathrm{sc}$ and $\mathrm{nd}$ are standard Jacobi functions~\cite{abramowitz1972handbook}]. The boundary condition can be written as 
\begin{align}
    2K(p_{\mathrm{mbb}})\frac{\mathrm{sc}\left( \frac{K(p_{\mathrm{mbb}})}{\delta},p_{\mathrm{mbb}}\right)}{\mathrm{nc}\left( \frac{K(p_{\mathrm{mbb}})}{\delta},p_{\mathrm{mbb}}\right)}=\frac{M|c|\delta L}{\mathrm{dc}\left( \frac{K(p_{\mathrm{mbb}})}{\delta},p_{\mathrm{mbb}}\right)}.
    \label{eq:bound_condition_mbb}
\end{align}
Equations~(\ref{eq:norm_condition_mbb}) and~(\ref{eq:bound_condition_mbb}) can be satisfied only for $N\leq N_{\mathrm{cr}}$, i.e., $f_{\mathrm{mbb}}$ can describe only such systems. The solution $f_{\mathrm{scatt}}$ describes systems with $N\geq N_{\mathrm{cr}}$. The calculation of $N_{\mathrm{cr}}$ will be given in the next section, see Eq.~\eqref{eq: DropCon}.

The subscripts `mbb' (many-body bound) and `scatt' (scattering) are motivated by the observation that for a large system
($L\to \infty$) $N_{\mathrm{cr}}$ is the maximal number of bosons that can be trapped by the impurity, see Ref.~\cite{Gunn, Kolomeisky2004} and the discussion in the next section.  
Note also that the chemical potential for the first (second) solution is negative (positive) for large system sizes since $p_{\mathrm{mbb}}\to1$ and $p_{\mathrm{scatt}}\to1$ for $L\to\infty$. This is another indication that the first solution describes a many-body bound state while the second is applicable if the bosons occupy scattering states.

\begin{figure}[t]
    \centering    
    \begin{subfigure}{0.5\textwidth}
    \centering
    \includegraphics[width=1\linewidth]{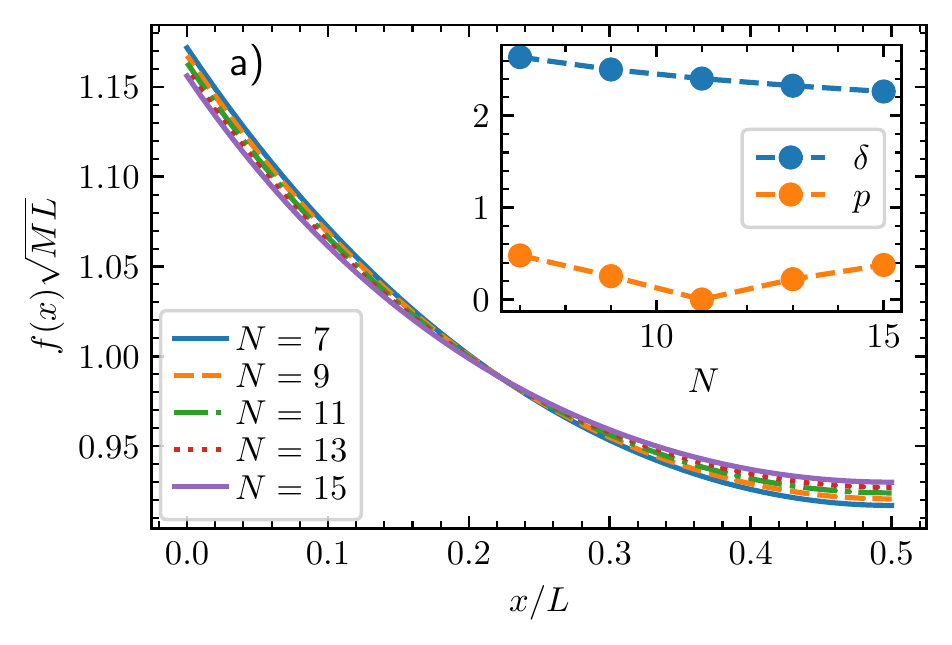}
    \end{subfigure}%
    \begin{subfigure}{0.5\textwidth}
    \centering
    \includegraphics[width=1\linewidth]{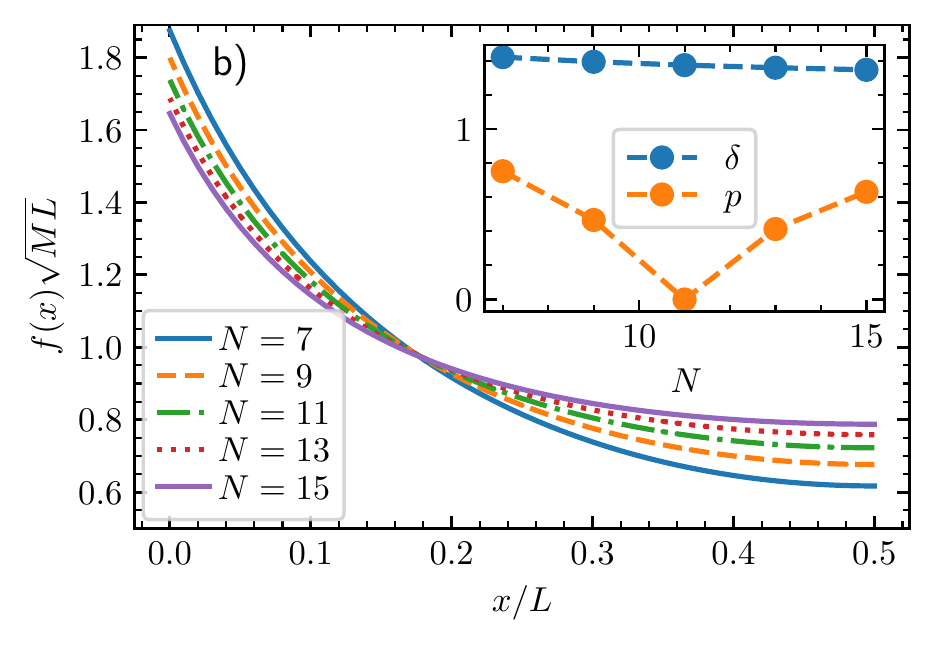}
    \end{subfigure}
    \begin{subfigure}{0.5\textwidth}
    \centering
    \includegraphics[width=1\linewidth]{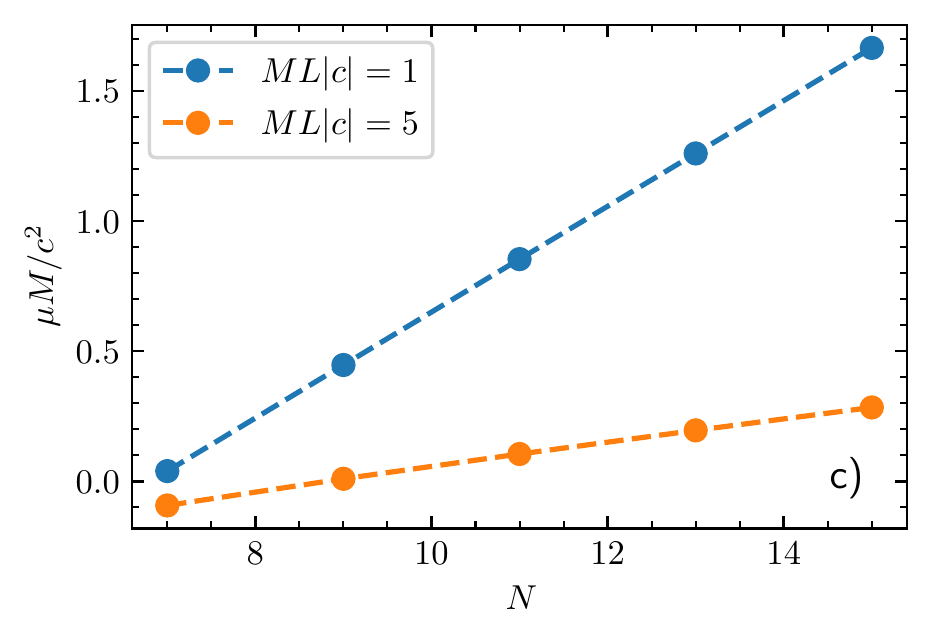}
    \end{subfigure}%
    \begin{subfigure}{0.5\textwidth}
    \centering
    \includegraphics[width=1\linewidth]{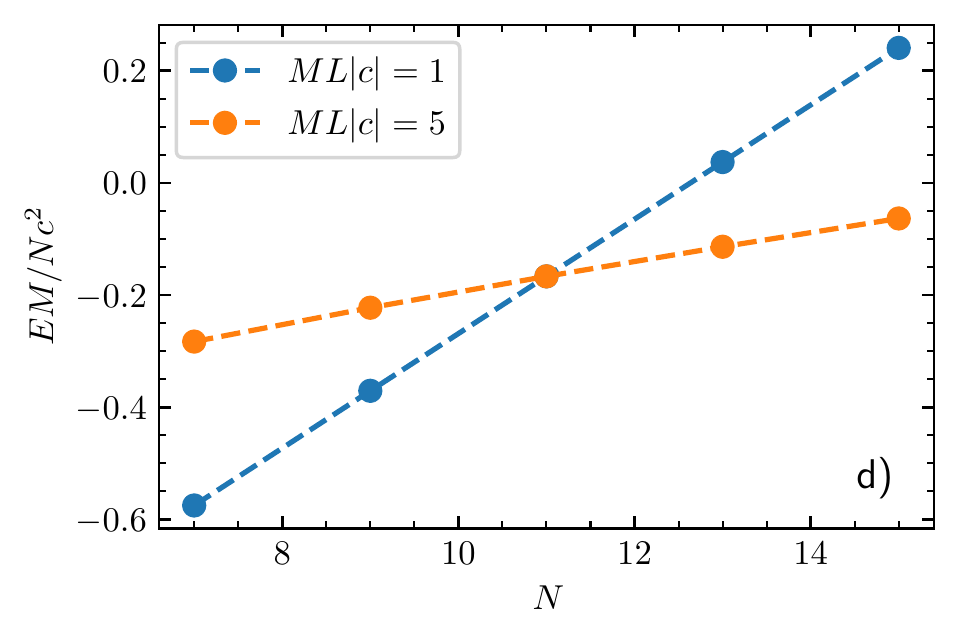}
    \end{subfigure}
    \caption{Upper panels: Mean-field solutions for different numbers of particles. Panel a) is for $LM|c|=1$; panel b) is for $LM|c|=5$. The relative interaction strength is fixed to $\alpha=-5$. Therefore the 
    maximal possible particle number in a bound state is $N_\mathrm{cr}=11$, see Eq.~(\ref{eq: DropCon}). The insets show the parameters $p$ and $\delta$ as a function of the particle number. The lower panels depict the corresponding chemical potential c) and the energy per particle d) for different system sizes (see legends).}
    \label{fig:MFSolutions}
\end{figure}

We illustrate mean-field solutions in Fig.~\ref{fig:MFSolutions} for different values of $N$ and $L$. At the position of the impurity, any solution $f$ reaches its maximum as a result of the attractive impurity-boson interaction.  Increasing the number of particles decreases the binding energy per particle (increases the energy per particle) due to the repulsive boson-boson interaction, see Fig.~\ref{fig:MFSolutions}~d), which also leads to a more flat profile of the density for the largest considered systems. The insets in panels~a) and b) show the parameters $p$ and $\delta$ as a function of the particle number. As it can be seen $\delta \to 1$ for increasing particle number. The parameter $p$ first drops down to $0$ at the critical particle number $N_\mathrm{cr}=11$, see Eq.~\eqref{eq: DropCon}, and then rises again towards~$1$ [Note that for $N\leq 11$ ($N>11$), $p$ corresponds to $p_{\mathrm{mbb}}$ ($p_{\mathrm{scatt}}$).]. For the largest ring size $p$ is larger; for $L\to\infty$ (not shown) we find empirically that $p\to1$ except in the vicinity of $N_\mathrm{cr}$. The chemical potential becomes negative for the largest ring size and $N\lesssim 11$ (Fig.~\ref{fig:MFSolutions}~c)), which is an indication of a bound state. We discuss this behavior in more detail in the following sections.

\subsection{Point of Transition}

The point of transition from one solution to another occurs at $p_{\mathrm{mbb}}=p_{\mathrm{scatt}}=0$ (cf. the insets in Figs.~\ref{fig:MFSolutions} a) and b)). In this case, the functions in Eqs.~(\ref{eq:MFmb}) and~(\ref{eq:MFscatt}) coincide: 
\begin{equation}
    f_{\mathrm{PoT}}(x)=\sqrt{\frac{\pi^2}{M gL^2\delta^2(N-1)}}\frac{1}{\cos\left(\frac{\pi (x -L/2)}{\delta L}\right)}\,.
    \label{eq: MFp0}
\end{equation}
The corresponding chemical potential reads
\begin{equation*}
    \mu_\mathrm{PoT}=\frac{\pi^2}{2M\delta^2L^2}.
\end{equation*}

It vanishes for large system sizes, i.e., $\mu_{\mathrm{PoT}}=0$ for $L\to\infty$. Normalization, and the boundary condition due to the delta-function potential determine the parameter~$\delta$
\begin{equation}
    \pi\tan\left(\frac{\pi}{2\delta}\right)=M|c|L\delta,
    \label{eq:delta_POT}
\end{equation}
and the critical number of bosons
\begin{equation}
    N_\mathrm{cr}=\frac{2|c|}{g}+1\,.
    \label{eq: DropCon}
\end{equation} 
Note that $N_{\mathrm{cr}}\to\infty$ when $g\to 0$, and $N_{\mathrm{cr}}=1$ when $g\to\infty$, in agreement with our discussion in Sec.~\ref{subsec:phys_int}.

Equation~(\ref{eq:delta_POT}) shows that $\delta$ is determined only by the dimensionless parameter $M|c|L$, whereas $N_{\mathrm{cr}}$ depends only on the ratio $|c|/g$. 
This decoupling of $\delta$ and $N_\mathrm{cr}$ is unexpected for systems with finite values of $L$. It suggests scale invariance of the problem at the point of transition, and leads to a number of surprising consequences. In particular, at the point of transition, the energy of the system also does not depend on~$L$:
\begin{equation}
    E=-\frac{N_\mathrm{cr}c^2M}{6}\,,
    \label{eq: Energycrit}
\end{equation}
which implies a state of zero pressure, in a sense that {\it it costs no energy to adiabatically change the radius of the ring}. This unique signature will later be used to identify the transition point in our numerical simulations. Note that Eq.~(\ref{eq: Energycrit}) provides a variational upper bound on the exact value of the energy. It rigorously  shows that more than a single boson can be trapped by the impurity, since this upper bound is below the ground-state energy of a single boson, $-Mc^2/2$, for $N_{\mathrm{cr}}\geq 3$.

\section{Zero-Density Limit: Mean-Field Results}
\label{chap:Expand}

In this section, we discuss the limit of vanishing density ($\rho\to0$) that occurs for a fixed number of bosons in a large system, i.e., $L\to\infty$ (see also Refs.~\cite{Kolomeisky2004,Gunn} and \ref{App:DetailsZeroDensity}).
This limit provides insight into the general results of the previous section.

\subsection{Many-body bound state}

The limit $\rho\to0$ leads to $p_{\mathrm{mbb}}\to1$ (cf. Fig.~\ref{fig:MFSolutions} and its discussion), so that Eq.~\eqref{eq:MFmb} can be written as  
\begin{equation}
    f_{\mathrm{mbb}}(x>0)=\sqrt{\frac{2\zeta(2\zeta+1)}{x_{\mathrm{mbb}}}}\frac{1}{(2\zeta+1)e^{x/x_{\mathrm{mbb}}}-e^{-x/x_{\mathrm{mbb}}}},
    \label{eq: MfmbLinfty}
\end{equation}
where 
\begin{equation*}\zeta=\frac{N_\mathrm{cr}-N}{N-1}, \qquad x_{\mathrm{mbb}}=\frac{1}{M|c|}\frac{N_{\mathrm{cr}}-1}{N_{\mathrm{cr}}-N}\; \left(=\frac{1}{M|c|}\frac{\zeta +1}{\zeta}\right).
\end{equation*}

The quantity $x_{\mathrm{mbb}}$ sets the characteristic width of the state\footnote{Note that $x_{\mathrm{mbb}}$ is proportional to the characteristic size of a one-body bound state $1/M|c|$. However, it can be much larger since $x_{\mathrm{mbb}}$ grows with $N$.}. If we define $\zeta=0$ for $N>N_\mathrm{cr}$, $\zeta$ can be seen as the `order' parameter for our system. 
Indeed, $\zeta$ is positive for a many-body bound state, and vanishes as we approach the point of transition. 
The respective chemical potential reads 
\begin{equation}
    \mu_{\mathrm{mbb}}=-\frac{1}{2Mx_{\mathrm{mbb}}^2}\,.
\end{equation}
It is negative which means that adding an additional boson lowers the total energy -- this is a typical characteristic of a bound state.  
The energy of the system is given by
 \begin{equation}
 \label{eq:EnergyBoundLinfty}
     E_{\mathrm{mbb}}= -\frac{N}{Mx_{\mathrm{mbb}}^2\zeta^2}\left(\frac{\zeta(\zeta+1)}{2}+\frac{1}{6}\right).
\end{equation}

Additionally, for large values of $|x|$ Eq.~(\ref{eq: MfmbLinfty}) yields
\begin{equation}
        f_{\mathrm{mbb}}(|x|\gg x_{\mathrm{mbb}})\simeq \sqrt{\frac{2\zeta}{(2\zeta+1)x_{\mathrm{mbb}}}} e^{-\frac{|x|}{x_{\mathrm{mbb}}}},
        \label{eq:mbb_tail}
\end{equation}
which corresponds to a typical tail of a bound state whose extension is defined by $x_{\mathrm{mbb}}$ (see, e.g., Ref.~\cite{landau1977quantum}). Note that Eq.~(\ref{eq:mbb_tail}) is valid only for $N<N_{\mathrm{cr}}$. At the point of transition ($\zeta\to 0, x_\mathrm{mbb}\to\infty$), another function will describe the tail of the state, see below.

\subsection{Point of Transition}

At $\zeta=0$, the mean-field solution of Eq.~\eqref{eq: MfmbLinfty} can be further simplified
\begin{equation}
    f_{\mathrm{PoT}}(x)=\sqrt{\frac{|c|M }{2}}\frac{1}{M |c x|+1}\,.
    \label{eq:MfCritLlarge}
\end{equation}
We see that for our many-body problem there is a finite probability to find a boson next to the impurity even at the threshold of binding. This clearly distinguishes the many-body problem from the one-body system [see Eq.~(\ref{eq:effective_physical})] where this probability vanishes in the limit $L\to\infty$. 

The characteristic length $x_{\mathrm{mbb}}$ diverges, and we need another quantity to describe the size of the state. It cannot be a root-mean-square radius, because of the $1/x$ tail of $f_{\mathrm{PoT}}(x)$. Still, we can define a meaningful size of the state as
\begin{equation*}
    x_\mathrm{PoT}=\frac{1}{|c|M}.
    \label{eq:WidthPoT}
\end{equation*}

This quantity defines the spatial region which contains half of the probability density, i.e., $\int_{-x_\mathrm{PoT}}^{x_\mathrm{PoT}}f_{\mathrm{PoT}}^2(x)\mathrm{d}x=1/2$. 
Note that $x_\mathrm{PoT}$ is given by the size of a one-particle bound state [Eq.~(\ref{eq:effective_physical})], which supports the physical picture provided in~Sec.~\ref{subsec:phys_int}. 

The energy of the system was already given in Eq.~\eqref{eq: Energycrit}, which is independent of the system size $L$. The chemical potential is zero. This implies that if we add more bosons, they must occupy scattering states. Hence, Eq.~(\ref{eq: Energycrit}) defines the energy for all systems with $N\geq N_{\mathrm{cr}}$ and $\rho\to0$.

\subsection{Scattering state}
\label{subsec:sc_state}

The function $f_{\mathrm{scatt}}$ of Eq.~\eqref{eq:MFscatt} in the limit $\rho\to0$ ($p_{\mathrm{scatt}}\to1$) can be approximated as follows
\begin{equation*}
        f_{\mathrm{scatt}}(x)\simeq \sqrt{\frac{1}{Mg L^2\delta^2(N-1)}}\ln\left(\frac{16}{1-p_{\mathrm{mbb}}}\right)\coth\left(\ln\left(\frac{16}{1-p_{\mathrm{mbb}}}\right)\left[\frac{x}{\delta L}+\frac{\delta-1}{2\delta}\right]\right).
\end{equation*}

For this solution, it is not possible to fulfil simultaneously normalization, and the delta-potential boundary conditions. If we impose the latter condition, we derive a non-normalizable wave function for $L\to\infty$. If we demand a normalized state, the resulting function is constant in space and does not include the impurity potential. We interpret this result as if all bosons occupy scattering states and distribute over the whole system until they are no longer affected by the impurity. This state is physical, however it is not the ground state, in which $N_\mathrm{cr}$ bosons are bound to the impurity, and other bosons occupy scattering states. The GPE cannot describe this physics because it assumes that all bosons occupy the same orbital. Hence, as soon as there are more bosons in the system than can be supported by the many-body bound state, all occupy scattering states within the mean-field approximation.

An appropriate variational ansatz for large (but finite) $L$ should include two parts, where the first part describes a many-body bound state, and the second one accounts for the bosons that occupy scattering states. In the low-density limit this leads to a Tonks-Girardeau gas formed outside the many-body bound state. We leave an investigation of such an ansatz for a future study.

\section{IM-SRG Results: Approach to the Zero-Density Limit}
\label{Chap:IMSRGZeroDensity}

To test our findings from the previous sections, we use the IM-SRG and ML-MCTDHX methods, which are briefly discussed in~\ref{App:NumMethods}, see the references given there for more details. 
Both numerical methods are able to capture corrections stemming from quantum fluctuations, and agree for the considered parameters.
Therefore, in this section we illustrate our numerical results only for IM-SRG, see~\ref{App:ComparisonSimos} for some ML-MCTDHX results.
 We focus on the question of approaching the limit $L\to\infty$. This allows us not only to test the mean-field  predictions but to also address  beyond-mean-field corrections that must be important far from the impurity, see Fig.~\ref{fig:OneImpurity:SketchSystem}~b).

\subsection{Energies}

\begin{figure}[t]
    \centering    
    \begin{subfigure}{0.5\textwidth}
    \centering
    \includegraphics[width=1\linewidth]{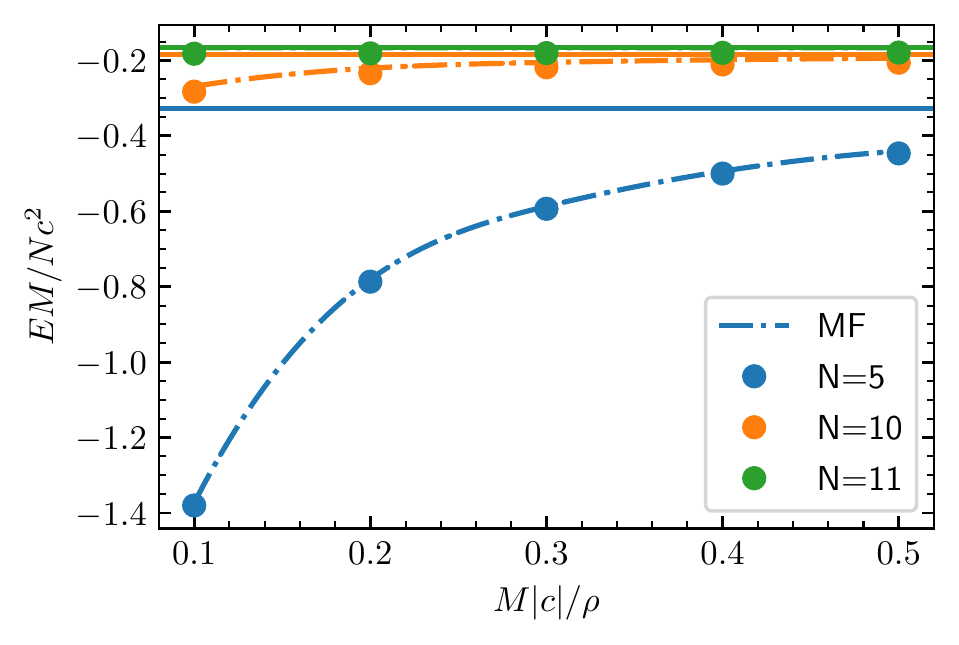}
    \end{subfigure}%
    \begin{subfigure}{0.5\textwidth}
    \centering
    \includegraphics[width=1\linewidth]{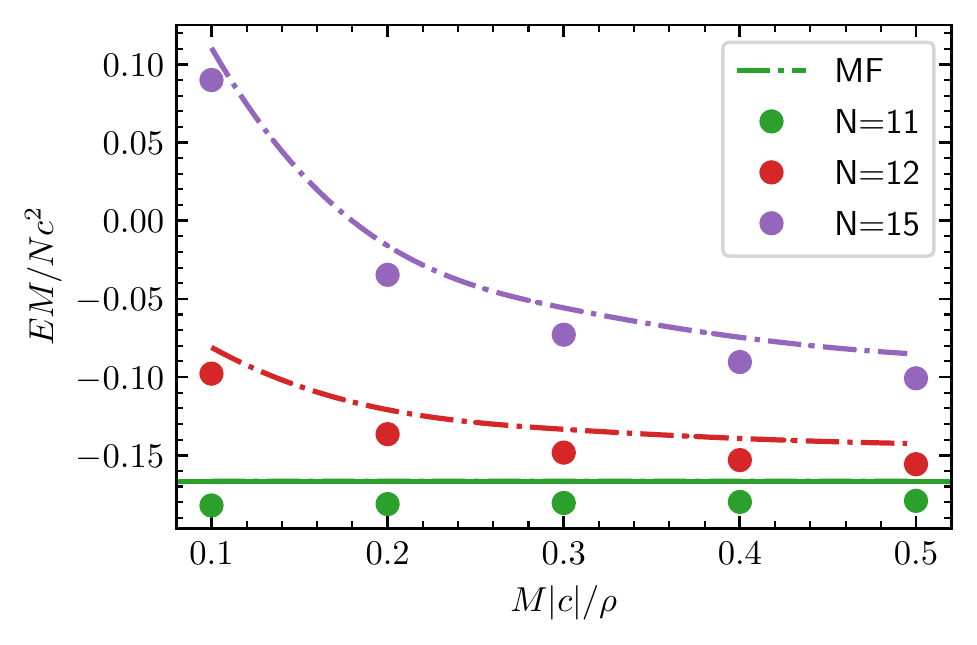}
    \end{subfigure}
    \caption{The energy per particle as a function of the inverse density $M|c|/\rho$ for different particle numbers. The circles are calculated using IM-SRG. The dashed-dotted curves represent our mean-field results. The solid (horizontal) lines correspond to the mean-field prediction for $\rho=0$, see Eq.~\eqref{eq:EnergyBoundLinfty}. The left panel shows the data for $N\leq N_{\mathrm{cr}}$. The right panel is for $N\geq N_{\mathrm{cr}}$. We fix $\alpha=-5$ which leads to $N_\mathrm{cr}=11$. Notice that the numerical error bars on the IM-SRG data calculated according to~\ref{App:NumMethods} are smaller than the sizes of the markers.  }
    \label{fig:LtoInfty:Energy}
\end{figure}

In Fig.~\ref{fig:LtoInfty:Energy}, we present the energy per particle as a function of the inverse density ($1/\rho$) for different values of $N$. For the considered parameters, the critical number of bosons that can be trapped by the impurity is $N_\mathrm{cr}=11$. 
We see that the IM-SRG data agree with the mean-field results well. Only small deviations are visible for $N\gtrsim N_{\mathrm{cr}}$. We attribute these deviations to residual beyond-mean-field effects naturally captured by the IM-SRG.
The energies of the systems with $N\geq N_\mathrm{cr}$ decrease and in the limit of $L\to\infty$, we expect them to approach the critical energy of Eq.~\eqref{eq: Energycrit}. Unfortunately, we cannot follow this convergence for larger values of $L$; we observed that the IM-SRG method is not accurate for $M|c|/\rho\geq 1$. 
In particular, for the largest considered particle numbers, the truncated flow equations diverge. This can be interpreted as a sign that the system becomes progressively more correlated, and IM-SRG cannot map the reference state (in our case a condensate) onto the real ground state of the system. 

For $N<N_{\mathrm{cr}}$, Fig.~\ref{fig:LtoInfty:Energy} shows that the energy increases with the size of the ring. This is a typical behavior for bound states (at least for $L\to\infty$), where the potential energy dominates the kinetic one. For $N=1$, this increase can be understood using the equation for the binding energy
\begin{equation*}
    \frac{\sqrt{2M|E|}}{M|c|}\mathrm{tanh}(\sqrt{2|E|M}L)=1,
\end{equation*}
which leads to $E=-\frac{M c^2}{2}-2Mc^2e^{-2M|c|L}$ for $M|c|L\gg 1$.
For $N=N_{\mathrm{cr}}$ the energy remains nearly constant with respect to the system size as predicted by Eq.~\eqref{eq: Energycrit}. There is a very weak dependence on $L$ pointing to beyond-mean-field effects. 
For $N>N_{\mathrm{cr}}$, the energy is a decreasing function of $L$. Our interpretation is that some bosons are now dropped out of the many-body bound state. Their kinetic energy decreases approximately as $1/L^2$, allowing us to conjecture the following behavior of the energy in the limit $L\to\infty$ 
\begin{equation}
    E\simeq -\frac{M N_{\mathrm{cr}}c^2}{6}+\frac{2\pi^2\mathcal{N}(\mathcal{N}+1)(2\mathcal{N}+1)}{3 M L^2},
    \label{eq:conjecture_odd}
\end{equation}
where $\mathcal{N}=(N-N_{\mathrm{cr}}-1)/2$ for odd values of $N-N_\mathrm{cr}$ and
\begin{equation}
    E\simeq -\frac{M N_{\mathrm{cr}}c^2}{6}+\frac{2\pi^2\mathcal{N}(\mathcal{N}+1)(2\mathcal{N}+1)}{3 M L^2}-2\frac{\mathcal{N}^2\pi^2}{ML^2},
    \label{eq:conjecture_even}
\end{equation}
with $\mathcal{N}=(N-N_{\mathrm{cr}})/2$ for even values of $N-N_\mathrm{cr}$. These expressions are the sums of the energy of the many-body bound state and the energy of the Tonks-Girardeau gas made of $N-N_{\mathrm{cr}}$ particles, assuming that there is no interaction between the bound state and bosons in scattering states\footnote{ Equations~(\ref{eq:conjecture_odd}) and (\ref{eq:conjecture_even}) assume that the size of the ring is sufficiently large in the sense that $1/k_F\gg1/|c|M$, where $k_F$ is the Fermi wavelength corresponding to the Tonks-Girardeau gas and $1/|c|M$ is the characteristic width of the many-body bound state. Assuming that $k_F=\pi\rho$, we derive the condition $M|c|/\rho\gg \pi$. This condition implies that the ring sizes used in Fig.~\ref{fig:LtoInfty:Energy} are too small to numerically confirm Eqs.~(\ref{eq:conjecture_odd}) and (\ref{eq:conjecture_even}).}.

All in all, the IM-SRG data support the existence of different physical scenarios that correspond to bound and scattering states. However, note that our numerical analysis cannot rule out the possibility that $N_{\mathrm{cr}}$ becomes larger when $L\to\infty$. In particular, we cannot rule out bound states with an infinite number of particles that are exponentially weakly bound in the limit $L\to\infty$. However, one does not expect this to happen because far from the impurity the bosons interact strongly (fermionize).

\begin{figure}[t]
\centering
    \begin{subfigure}{0.5\textwidth}
    \centering
    \includegraphics[width=1\linewidth]{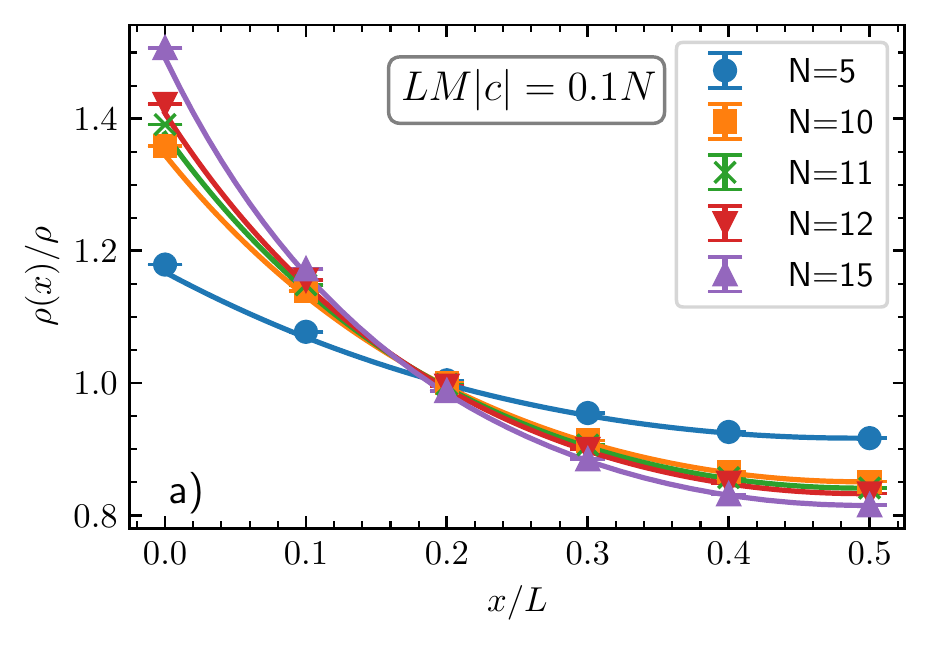}
    \end{subfigure}%
    \begin{subfigure}{0.5\textwidth}
    \centering
    \includegraphics[width=1\linewidth]{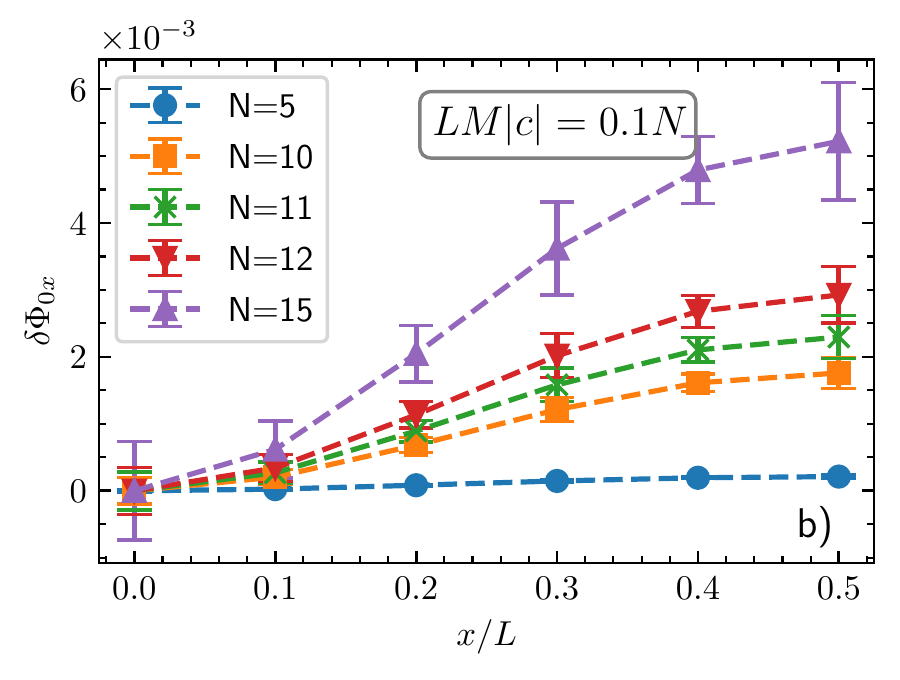}
    \end{subfigure}
    \begin{subfigure}{0.5\textwidth}
    \centering
    \includegraphics[width=1\linewidth]{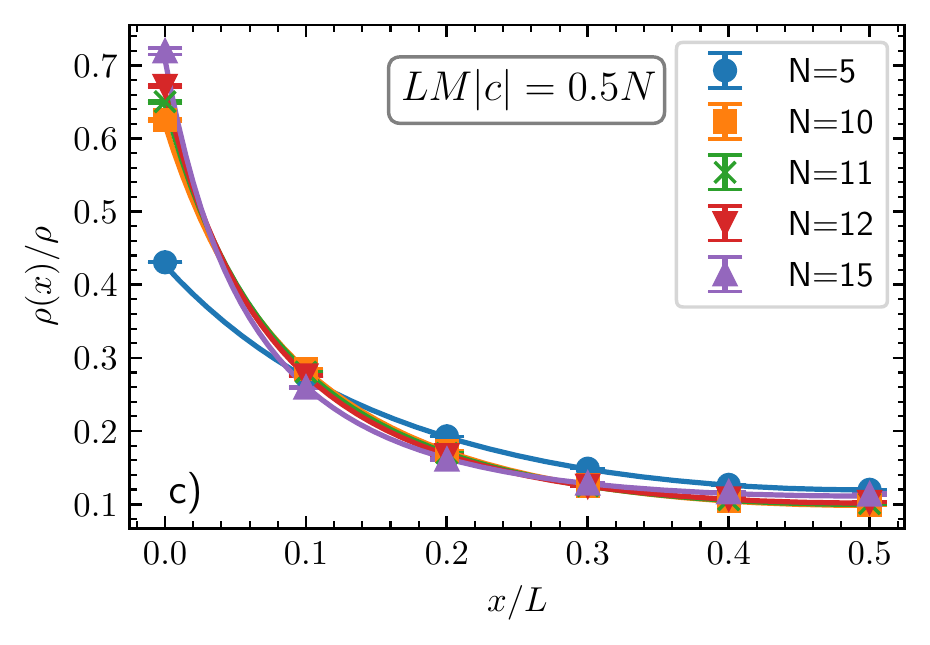}
    \end{subfigure}%
    \begin{subfigure}{0.5\textwidth}
    \centering
    \includegraphics[width=1\linewidth]{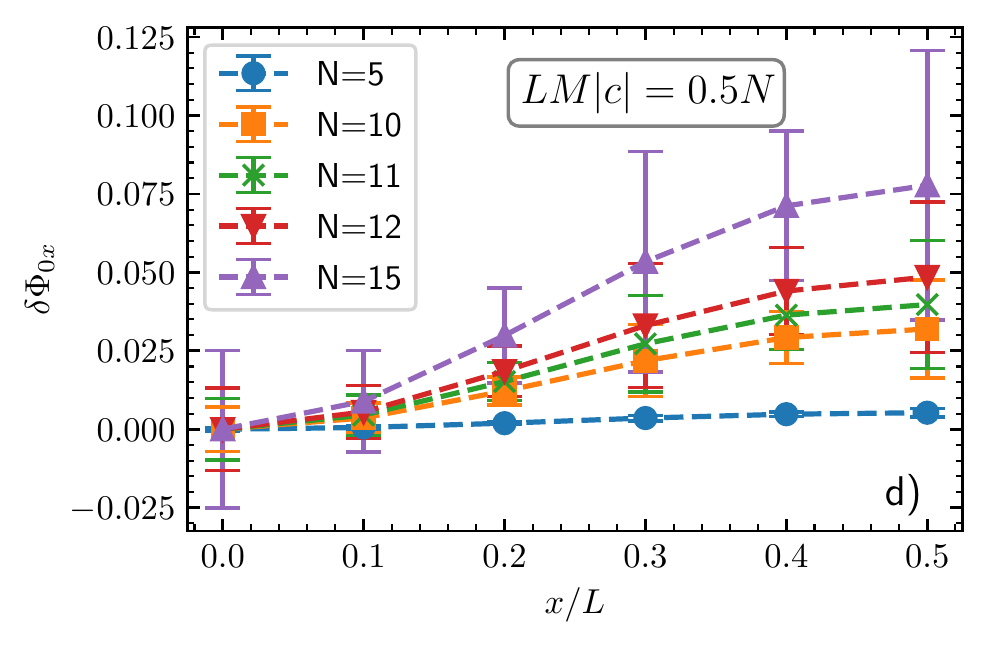}
    \end{subfigure}
\caption{Panels a) and c) show the density of the Bose gas. Circles are calculated with IM-SRG, and the solid curves are the corresponding mean-field results (Eqs.~\eqref{eq:MFmb},~\eqref{eq: MFp0}). Panels b) and~d) demonstrate phase fluctuations whose non-zero values reveal presence of beyond-mean-field correlations. The dashed curves are plotted to guide the eye. The data show  many-body bound, critical and scattering states. We fix $\alpha=-5$, thus, the critical number of bosons supported by the bound state is $N_\mathrm{cr}=11$. Panels a) and b) are for systems with $LM|c|=0.1N$ ($M|c|/\rho=0.1$). Panels c) and d) refer to $LM|c|=0.5N$ ($M|c|/\rho=0.5$). The numerical error bars are calculated according to the prescription given in \ref{App:NumMethods}.}
\label{fig:Ltoinfty:phasefluc}
\end{figure}

\subsection{Densities and Phase Fluctuations}

Here, we calculate the density of the Bose gas
\begin{equation}
    \rho(x)=\braket{\Phi_{gr}|\sum\limits_{i=1}^N\delta({x}-x_i)|\Phi_{gr}},
    \label{eq:density}
\end{equation}
in the ground state, $\Phi_{gr}$. We also investigate beyond-mean-field effects. To this end, we estimate phase fluctuations (also known as phase correlations), $\delta\Phi_{xx'}$, from the one-body density matrix according to the prescription (see, e.g.,~\cite{Popov1983,Petrov2000,Pethick2002})
\begin{equation}
    \rho(x,x')\equiv \braket{\Phi_{gr}|\rho(x,x')|\Phi_{gr}}=\sqrt{\rho(x)\rho(x')}\exp\left\{-\frac{\delta\Phi_{xx'}}{2}\right\}.
    \label{eq:phase_def}
\end{equation}
The quantity $\delta\Phi_{xx'}$  is a measure of the off-diagonal quasi-long-range order, which vanishes for a condensate (mean-field) state. Note that $\delta\Phi_{xx'}$ is not only a convenient theoretical object for studying the importance of the beyond-mean-field effects. It also leads to experimental indicators of phase coherence that are observable through Bragg spectroscopy and interferometry, see, e.g., Ref.~\cite{Hellweg2001}. 

In Fig.~\ref{fig:Ltoinfty:phasefluc}, we show $\rho(x)$ and $\delta\Phi_{xx'}$ for $L=0.1 N/M|c|$ and $L=0.5 N/M|c|$. For all considered parameters, the IM-SRG and mean-field results agree on the density profile of the Bose gas. The density is the highest in the vicinity of the impurity, as expected. For the largest values of $N$, it features a weak dependence on $N$ irrespective of the (considered) ring size.
In spite of this, there is a noticeable increase of beyond-mean-field correlations as identified by the non-vanishing phase fluctuations. 
Their effect is more pronounced for the largest ring and $N\geq N_{\mathrm{cr}}$, especially in the region with low densities. This observation is in agreement with the physical picture outlined in Fig.~\ref{fig:OneImpurity:SketchSystem}: low densities lead to strong boson-boson interactions, which can be quantified by $Mg/\rho(x)$.  Surprisingly, IM-SRG and mean-field results are in a reasonable agreement even when $Mg/\rho(x)$ is of the order of unity, where the mean-field treatment is not expected to be valid. It is also worthwhile noting that the mean-field approximation is valid, in particular Eq.~\eqref{eq: DropCon}, even for the smallest
non-trivial bound system – a two-boson artificial atom, see~\ref{App:Few-body_Limit}.

Note that phase fluctuations are the strongest for the largest considered $N$. This can be rationalized in the following way. For $N=15$, a few bosons are not trapped by the impurity. Therefore, the probability of strong boson-boson interactions far away from the impurity is high leading to large phase fluctuations. In contrast, for small particle numbers (e.g., $N=5$), phase fluctuations may be small even if the density is low. The probability of finding two bosons outside the many-body bound state in this case is exponentially suppressed.

\section{Mobile Impurity}
\label{chap:mobile}

A single mobile impurity atom in a weakly interacting Bose gas is an experimentally relevant system~\cite{Catani2012,Widera2012}, which motivated various theoretical studies of a `Bose-polaron', see, for example, Refs.~\cite{Grusdt2017,QMCPolaron,PANOCHKO_2019,ArtemPolaron,Jager2020,mistakidis2021radiofrequency,mistakidis2019quench}. Here, we complement those studies by considering the many-body bound state that follows from our results in the previous sections.

\begin{figure}[t]
    \centering
     \begin{subfigure}{0.5\textwidth}
    \centering
    \includegraphics[width=1\linewidth]{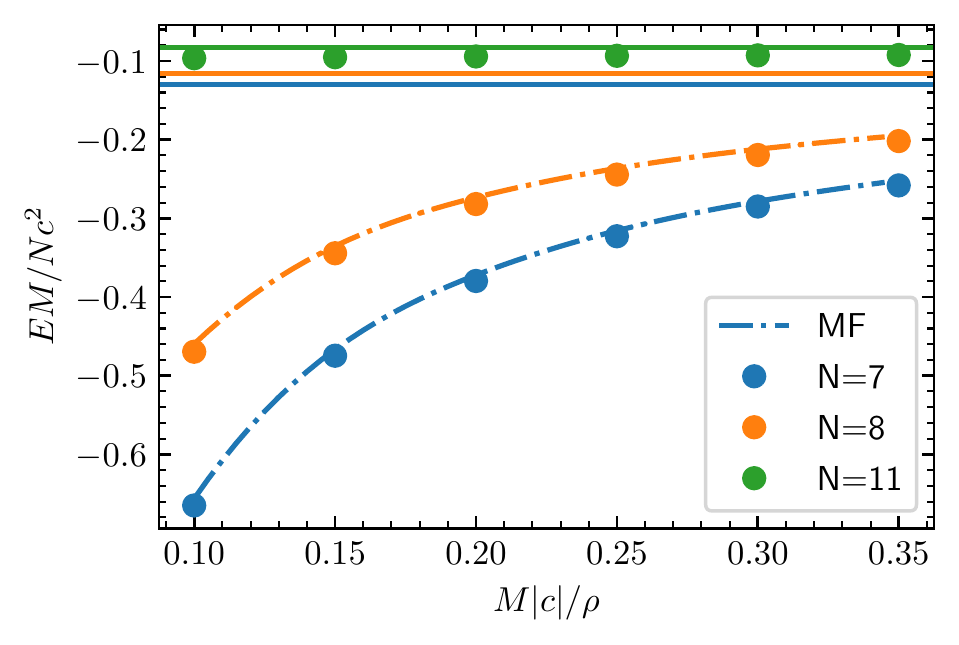}
    \end{subfigure}%
    \begin{subfigure}{0.5\textwidth}
    \centering
    \includegraphics[width=1\linewidth]{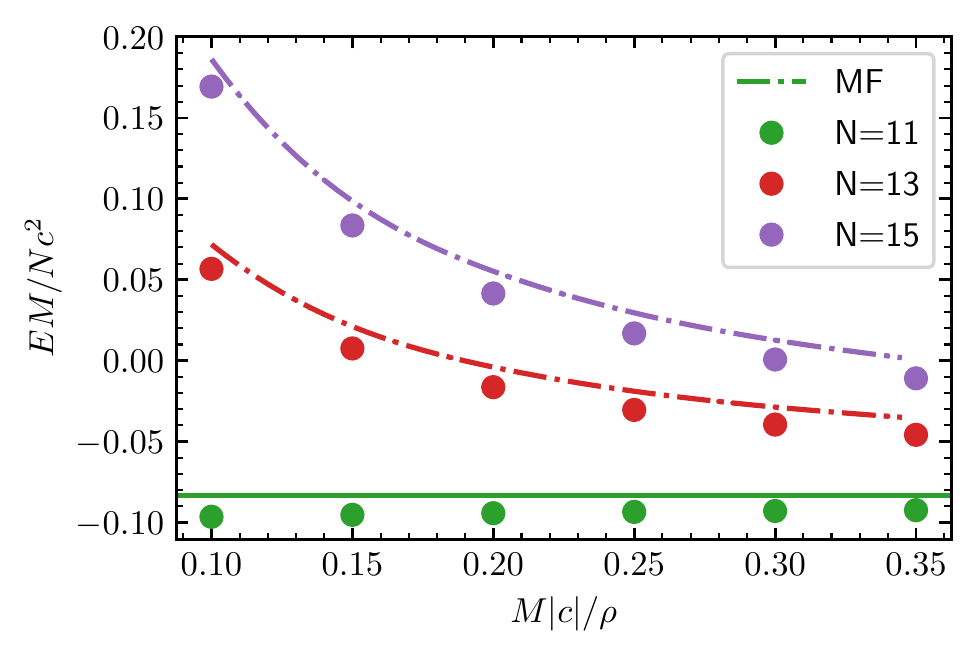}
    \end{subfigure}
    \caption{The energy per particle as a function of the inverse density $1/\rho$ for a mobile impurity ($m=M$) for different values of $N$. The circles are calculated using IM-SRG. The dashed-dotted curves represent the mean-field results. The solid (horizontal) lines correspond to the mean-field prediction for $\rho=0$, see Eq.~\eqref{eq:EnergyBoundLinfty}. The left (right) panel shows the energy for $N\leq N_{\mathrm{cr}}$ ($N\geq N_{\mathrm{cr}}$). We fix $\alpha=-5$ which leads to $N_\mathrm{cr}=11$. Notice that the numerical error bars calculated within IM-SRG according to~\ref{App:NumMethods} are smaller than the sizes of the markers. }
    \label{fig:Mobile:Energy}
\end{figure}

\subsection{Mean-field analysis}
To investigate an impurity with a finite mass, we use the mean-field ansatz in the frame `co-moving' with the impurity~\cite{GROSS_1962,ArtemPolaron,PANOCHKO_2019,Jager2020,Enss2020,Guenther2021}.  This frame is introduced via the set of new coordinates
\begin{equation}
    z_i=L\theta(y-x_i)+x_i-y,
    \label{eq: relCoord}
\end{equation} 
where $\theta(x)$ is the Heaviside step function. These coordinates allow us to exclude the position of the impurity from the Hamiltonian (similarly to the Lee-Low-Pines transformation~\cite{Lee1953})\footnote{Transformation to the `co-moving' frame allows us to use the analytical results of the previous sections. The mean-field approximation in the laboratory frame will lead to a system of coupled Gross-Pitaevskii equations, which one should solve numerically, see~\ref{App:ComparisonSimos}.}. In the new coordinates, the Hamiltonian reads as follows
\begin{equation}
    \mathcal{H}_P=-\frac{1}{2M}\sum_i^N\frac{\partial^2}{\partial z_i^2}-
    \frac{1}{2m}\left(\sum_i^N\frac{\partial}{\partial z_i}\right)^2+\frac{iP}{m}\sum_i^N\frac{\partial}{\partial z_i}+g\sum_{i<j} \delta(z_i-z_j)+c\sum_{i=1}^N\delta(z_i),
    \label{eq:LLP_HP}
\end{equation}
where $P$ is a quantum number -- the total (angular) momentum of the system. 
For simplicity, we consider the case $P=0$, which corresponds to the ground-state manifold. 

The Gross-Pitaevskii equation that follows from Eq.~(\ref{eq:LLP_HP}) reads (see, e.g.,~\cite{ArtemPolaron}):
\begin{equation}
    -\frac{1}{2\kappa}\frac{d^2f}{dz^2}+g(N-1)f(z)^3+c\delta(z)f(z)=\mu f(z)\,,
    \label{eq:LLP_GPE}
\end{equation}
where  $\kappa=mM/(m+M)$ is the reduced mass.
This equation is equivalent to Eq.~\eqref{eq:GPE} up to a change of the mass of the boson $M$ to $\kappa$. In this sense, all of our mean-field results from Secs.~\ref{chap:mf} and~\ref{chap:Expand} also apply to a mobile impurity.

\subsection{IM-SRG results}

We use IM-SRG to validate the mean-field predictions of Eq.~(\ref{eq:LLP_GPE}). We focus on the case with $m=M$. In Fig.~\ref{fig:Mobile:Energy}, we show the ground-state energy as a function of $1/\rho$ - similar to Fig.~\ref{fig:LtoInfty:Energy}. First of all, we see that the mean-field and IM-SRG results are in agreement. Furthermore, just like before, the energy of the system with $N<N_{\mathrm{cr}}$ increases as a function of $L$. For the critical number of bosons, the energy remains nearly constant. Note that according to Eq.~\eqref{eq: DropCon} the critical number of bosons does not depend on the mass of the impurity. Our numerical simulations confirm this result. Finally, we used IM-SRG to calculate the density and phase fluctuations of the Bose gas in the presence of a mobile impurity. The comparison of the mean-field predictions to the IM-SRG is similar to the one presented in Fig.~\ref{fig:Ltoinfty:phasefluc}. Therefore, we refrain from discussing it further.

\section{Summary \& Outlook}
\label{sec:concl}

We studied a one-dimensional artificial atom made of bosons. 
First, we analyzed this system within the mean-field approximation, and presented two possible solutions. In the limit $L\to\infty$, the solutions correspond to two different physical scenarios with the bosons bound (or not) to the impurity. The critical state in between these scenarios is a zero-pressure state, meaning that its energy does not depend on the radius of the ring. We presented analytical expressions that describe this state.

Second, we investigated the system numerically using beyond-mean-field methods (IM-SRG and ML-MCTDHX). 
Our numerical simulations justified the use of the mean-field approximation for studying artificial atoms from bosons in one dimension. They confirmed the existence of bound, critical and scattering states in the system. Despite the validity of the obtained mean-field solutions, we argued that quantum fluctuations are present in the tail of the wavefunctions. Therefore, only the bosons near the impurity are described with a mean-field ansatz well. Bosons far away from the impurity are strongly interacting, supporting the phenomenological argument of Ref.~\cite{TrappingCollapse}. However, their influence on the system can be neglected for particle numbers smaller than the critical one, because the attraction from the impurity assures a sufficiently large region with high density where particles are weakly interacting\footnote{Note that in~\ref{App:Few-body_Limit} we validate the mean-field solution even for a two-boson system where the strength of the boson-boson repulsion can be of the same order of magnitude as the one of the boson-impurity attraction.}. Although, we mainly focused on a heavy impurity, we also showed that our results are applicable for a mobile one. 

Further studies are needed to understand Bose systems with $N_{\mathrm{cr}}+1$ particles in the limit $L\to\infty$. Our results indicate that the mean-field approach is not suitable for such studies. In particular, it cannot be used to calculate the effective boson-artificial-atom interactions. The knowledge of this interaction will simplify the analysis of low-density Bose gases with attractive impurity-boson interactions. 

Our results pave the way for investigations of many-atom physics using artificial atoms as elementary building blocks. For example, a lattice of heavy impurities immersed in a Bose gas may feature different phases (e.g., Mott insulator and superfluid) depending on the strength of the boson-impurity and boson-boson interactions. Dilute systems of artificial atoms based upon mobile impurities can enjoy the physics of cold gases. To explore that context, one needs to understand the effective interaction between two artificial atoms.

\vspace{2em}

\noindent \textbf{Acknowledgements}: 
This work has received funding from the DFG Project No. 413495248 [VO 2437/1-1] (F. B., H.-W. H., A. G. V.)
and European Union's Horizon 2020 research and innovation programme under the Marie Sk{\'l}odowska-Curie Grant Agreement No. 754411 (A. G. V.). M. L. acknowledges support
by the European Research Council (ERC) Starting Grant No. 801770 (ANGULON). S. I. M. acknowledges support from the NSF through a grant for ITAMP at Harvard University.

\appendix

\section{Numerical Methods}
\label{App:NumMethods} 

In this Appendix, we briefly discuss the two numerical methods used in this work. The \textit{first method} is called the flow equation approach or IM-SRG (``in-medium similarity renormalization group''). Our numerical implementation of this approach is based upon previous works~\cite{TwoImpIMSRG,ArtemPolaron} (see also Ref.~\cite{Volosniev2017} for a study of the Lieb-Liniger gas), which 
are inspired by the methods known in condensed matter and nuclear physics (see e.g.~\cite{Kehrein2006, Tsukiyama2011, Hergert2016}).
The \textit{second method} is called the multi-layer multi-configuration time-dependent Hartree method for atomic mixtures (ML-MCTDHX)~\cite{cao2017unified} (see also a relevant review on the topic~\cite{Lode2020}). It is a variational approach that has been extensively used, among others, for studying systems with impurities~\cite{mistakidis2021radiofrequency,mistakidis2020pump,mistakidis2020induced,mistakidis2020many,mistakidis2019quench,Mistakidis2019}. 

\subsection{Flow Equation Approach (IM-SRG)}
The flow equation approach (block)-diagonalizes the Hamiltonian in second quantization,
\begin{equation}
    H=\sum\limits_{i,j}A_{i,j}a_i^\dagger a_j + \sum\limits_{i,j,k,l}B_{ijkl}a_i^\dagger a_j^\dagger a_k a_l,
\end{equation}
via the so-called flow equation
\begin{equation}
    \frac{dH}{ds}=[\eta, H].
    \label{eq:FlowEquation}
\end{equation}
Here, $s$ is the flow parameter, which formally plays a role of (imaginary) time. The generator of the flow $\eta$ has to be chosen such that the off-diagonal matrix elements vanish in the limit $s\to\infty$~\cite{Kehrein2006}. 

In this work, we aim to decouple the ground state from the rest of the Hilbert space. Therefore, we normal order the Hamiltonian using a reference state following the prescription in Ref.~\cite{Volosniev2017}. This leads to the normal-ordered Hamiltonian
\begin{equation}
    H=E\mathbb{I}+\sum\limits_{i,j}f_{i,j}:a_i^\dagger a_j: + \sum\limits_{i,j,k,l}\Gamma_{ijkl}:a_i^\dagger a_j^\dagger a_k a_l:,
\end{equation}
where we denote normal ordered operators with $:O:$. The matrix elements $f_{ij}$ and $\Gamma_{ijkl}$ describe one- and two-particle excitations from the reference state. For the generator, we use
\begin{equation}
    \eta(s)=f_{i0}(s):a_i^\dagger a_j:+\Gamma_{ij00}(s):a_i^\dagger a_j^\dagger a_k a_l: -\mathrm{h.c.},
\end{equation}
these are the matrix elements which need to vanish in order to decouple the ground state from the excitations. Therefore, once the flow equation converges, our ground state is decoupled.

The transformation governed by the flow equations can also be understood as a mapping of the reference state onto the real ground state of the system. Since we are interested in a system of bosons, it is reasonable to use condensate as reference state. Our reference state is constructed iteratively: Starting from the ground state solution of the non-interacting Hamiltonian, the density is calculated and used as the new reference state. This procedure is repeated until the density converges. Note that also other choices for the reference state are possible such as the mean-field solution, see Ref.~\cite{TwoImpIMSRG}. For our system of interest such a reference state leads to the same result.

 Induced higher order terms make it impossible to solve Eq.~\eqref{eq:FlowEquation} exactly, and should be truncated. In our truncation scheme, we truncate at the two-body level, while keeping three-body operators which contain at least one $a_0^\dagger a_0$ operator. This leaves us only with zero-, one- and two-body operators in Eq.~\eqref{eq:FlowEquation} which leads to a system of coupled, closed, non-linear differential equations, which we solve numerically~\cite{Volosniev2017, TwoImpIMSRG}. We estimate the error due to the neglected pieces (called $W$) using second order perturbation theory
\begin{equation}
    \delta E\simeq\sum\limits_p\frac{\left(\braket{\Phi_p|\int_0^\infty W(s)ds|\Phi_\mathrm{ref}}\right)}{\braket{\Phi_p|H|\Phi_p}-\braket{\Phi_\mathrm{ref}|H|\Phi_\mathrm{ref}}}\,,
\end{equation}
where $\Phi_p$ is a state that contains three-body excitations and $\Phi_\mathrm{ref}$ is our reference state.

To construct the Hamiltonian in second quantization we use the solution of the one-body Hamiltonian of our system. Since we can only work with a finite Hilbert space, we solve the flow equations for different numbers of basis states (in our case $n\in[11, 13, 15, 17, 19, 21]$). For the energy, we fit these values with
\begin{equation}
    E(n)=E(n\to\infty)+\frac{A}{n^\delta}
\end{equation}
to obtain the result in infinite Hilbert space. For other observables, such a fit is not always possible. In such cases, we take the result for the largest Hilbert space as our result and estimate the error by the largest deviation between the results for the different numbers of basis states. So there are in total two contributions to our error bars: The truncation error from neglecting higher order terms in the flow equation and the truncation error due to a finite Hilbert space. 

For a more detailed description of the method we refer to Ref.~\cite{Volosniev2017}, where the flow equations and our estimate of the truncation error are introduced, see also Ref.~\cite{TwoImpIMSRG} for information about calculation of observables and a detailed explanation of our estimate of error bars.

\subsection{ML-MCTDHX approach}

In the ML-MCTDHX approach, the Hilbert space is truncated in a variationally optimal manner. To this end, one employs a time-dependent moving basis in which the system is instantaneously optimally represented through time-dependent permanents~\footnote{For a multicomponent setting, the variational ansatz has a multilayer structure allowing one to include both intra- and interspecies correlations, see Ref.~\cite{cao2017unified}.  Here, we describe a reduction of ML-MCTDHX to a single-component system that is investigated.}. 
In this sense, the many-body wave function is expressed with respect to bosonic number states $\ket{\vec{n}}\equiv\ket{n_1,n_2,\dots,n_D;t}$ and time-dependent expansion coefficients $C_{\vec{n}}(t)$ as follows  
\begin{equation}
\ket{\Psi(t)}=\sum_{\vec{n}} C_{\vec{n}}(t) \ket{n_1,n_2,\dots,n_D;t}.\label{wfn_MCTDHB}    
\end{equation}
Here, $\ket{\vec{n};t}$ built upon time-dependent single-particle functions $\varphi_i(t)$ with $i=1,2,\dots,D$. The summation in Eq.~(\ref{wfn_MCTDHB}) is performed over all possible combinations $n_i$ such that the total number of bosons $N$ is conserved. 
In our numerical implementation, the single-particle functions $\varphi_i(t)$ are expanded in a time-independent primitive basis of dimension $\mathcal{M}$~\footnote{In the limit where $D=\mathcal{M}$ the wave function expansion of Eq.~(\ref{wfn_MCTDHB}) is equivalent to a full configuration interaction approach, while for $D=1$ it reduces to a  single product state, which automatically satisfies symmetrisation conditions for bosons, and thus corresponds to the mean-field approximation.} that is based upon a sine discrete variable representation for the box potential with hard-wall boundary conditions at $\pm L/2$. To calculate the ground-state wave function of the many-body setting, we determine the underlying equations-of-motion for the coefficients $C_{\vec{n}}(t)$ and the single-particle functions $\varphi_i(t)$ following the Dirac-Frenkel~\cite{frenkel1934wave} variational principle. 
An imaginary time propagation method is used to obtain the system's ground state configuration. 
More details on the ingredients of this variational method can be found in Refs.~\cite{cao2017unified,kronke2013non}.

\section{ML-MCTDHX results}
\label{App:ComparisonSimos}

\begin{figure}[t]
\centering
    \begin{subfigure}{0.5\textwidth}
    \centering
    \includegraphics[width=1\linewidth]{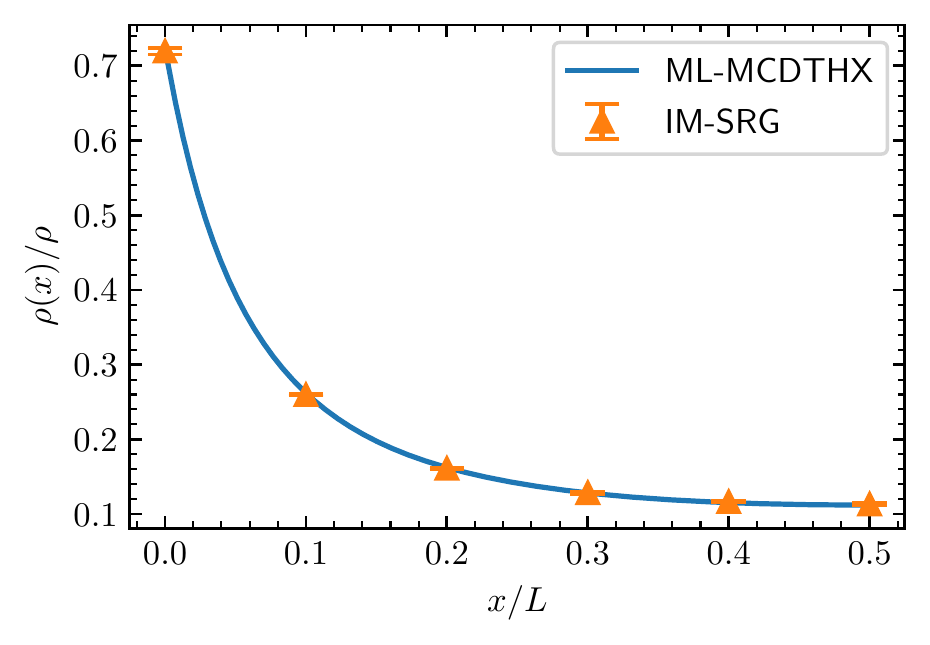}
    \end{subfigure}%
    \begin{subfigure}{0.5\textwidth}
    \centering
    \includegraphics[width=1\linewidth]{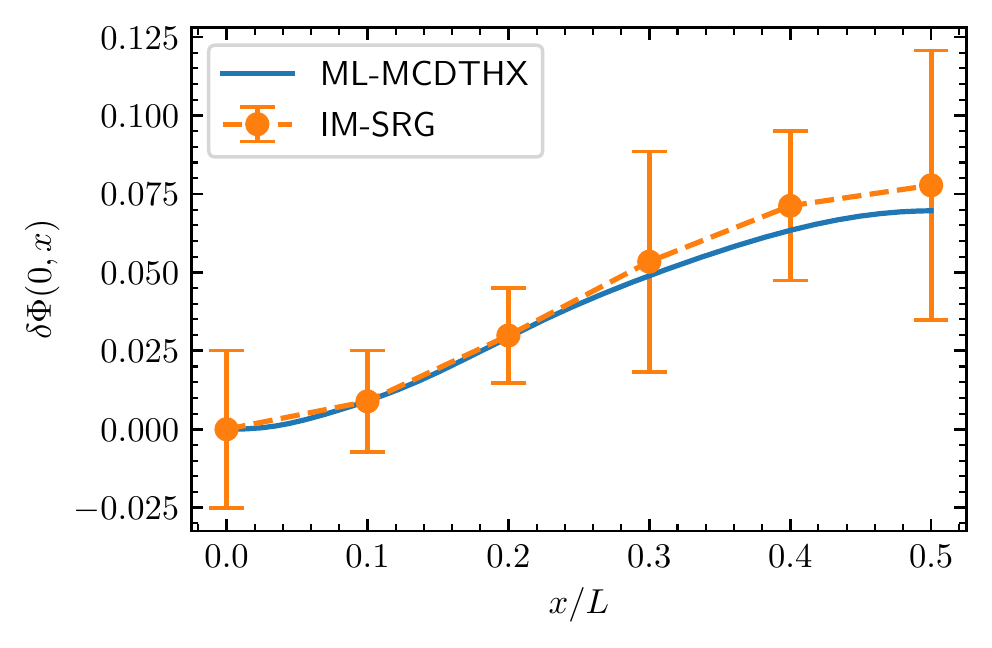}
    \end{subfigure}
\caption{Density (left) and phase fluctuations (right) for the Bose gas using periodic boundary conditions calulcated with IM-SRG (circles) and ML-MCTDHX (solid curves). The dashed curves are added to guide the eye. The parameters of the system are: $\alpha=-5$, $LM|c|=0.5N$, and $N=15$. In ML-MCDTHX, we used $D=5$ orbitals. The energy per particle in IM-SRG is $EM/Nc^2=-0.100722\pm 0.000089$ and in ML-MCDTHX it is $EM/Nc^2=-0.097$. The numerical error bars are calculated according to the prescription given in \ref{App:NumMethods}.}
\label{fig:App:ComparisonPerio}
\end{figure}

In the main text, the analytical solution for the bound state has been benchmarked against IM-SRG data. We have checked that these results are in agreement with the predictions of the well-established ML-MCDTHX approach. This is illustrated in Fig.~\ref{fig:App:ComparisonPerio} where the densities and phase fluctuations for the largest value of $N$ considered in the main text are shown (cf. Fig.~\ref{fig:Ltoinfty:phasefluc}).

Below, we study a system in a box potential, thus, exploring the formation of the artificial atom from bosons in the presence of hard-wall boundary conditions. This allows us to further understand the validity of the relatively novel IM-SRG method. 
Afterwards, we discuss the mean-field approximation to a mobile impurity in a Bose gas without the transformation to relative coordinates, Eq.~\eqref{eq: relCoord}.

\begin{figure}[t]
\centering
    \begin{subfigure}{0.5\textwidth}
    \centering
    \includegraphics[width=1\linewidth]{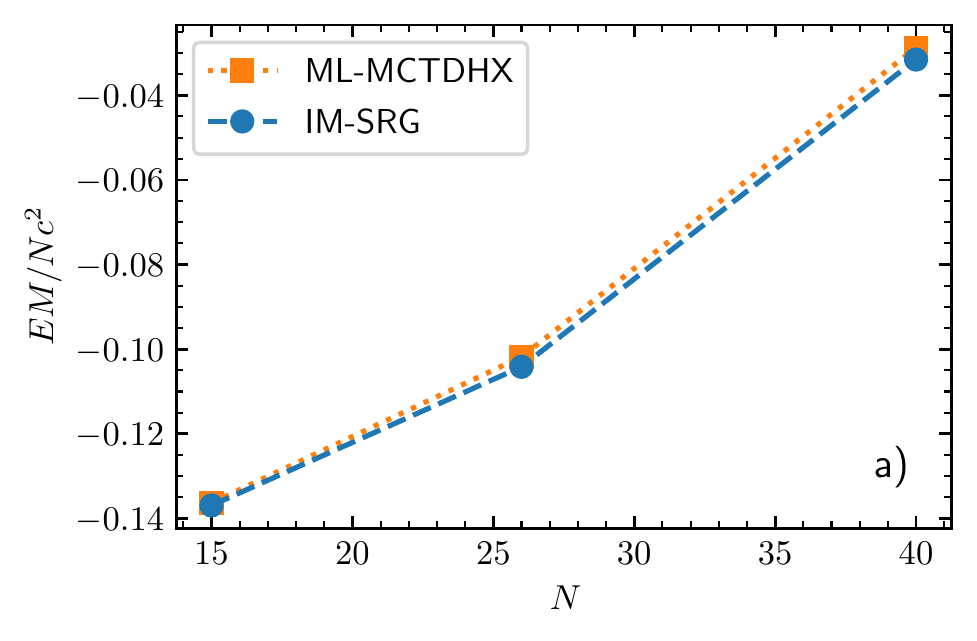}
    \end{subfigure}%
    \begin{subfigure}{0.5\textwidth}
    \centering
    \includegraphics[width=1\linewidth]{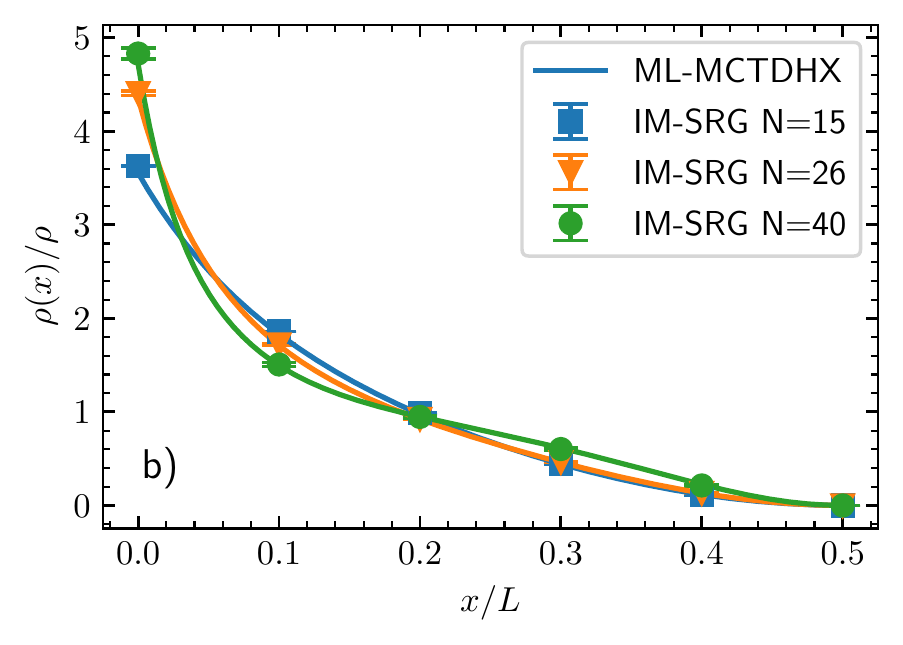}
    \end{subfigure}
    \begin{subfigure}{0.5\textwidth}
    \centering
    \includegraphics[width=1\linewidth]{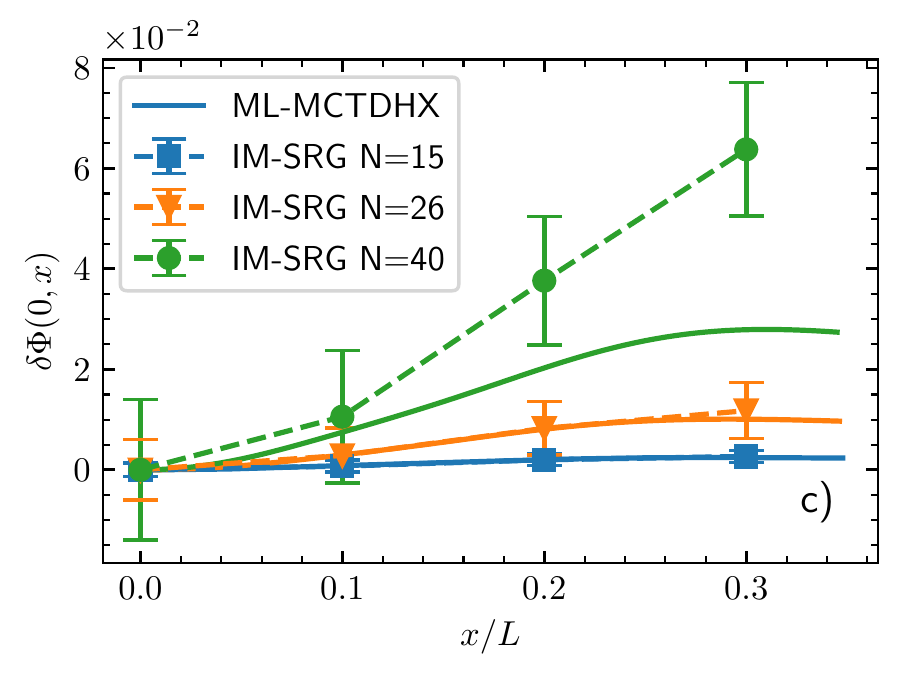}
    \end{subfigure}%
    \begin{subfigure}{0.5\textwidth}
    \centering
    \includegraphics[width=1\linewidth]{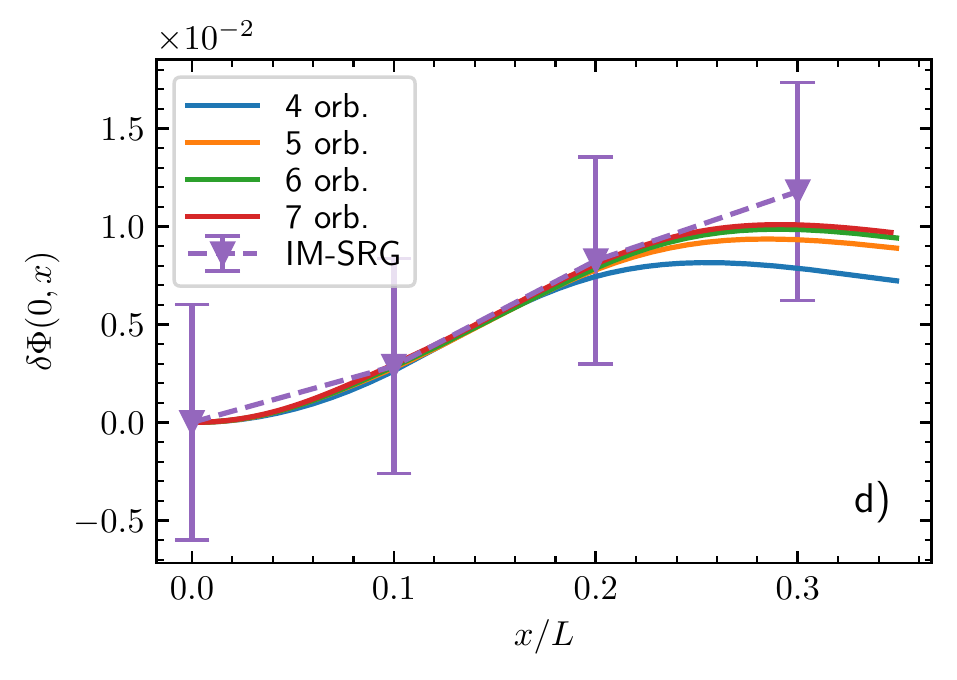}
    \end{subfigure}
\caption{Observables for a Bose gas in a box trap assuming that there is an impurity in the middle of the trap. The parameters of the system are: $\alpha=-12.5$, $LM|c|=0.25N$, and $N=15,\,26,\,40$. [Note that the critical particle number for these parameters in a ring would be $N_\mathrm{cr}=26$]. Panels a), b) and c) show correspondingly the energy, the density and phase fluctuations 
calculated with IM-SRG and ML-MCDTHX for different particle numbers. Solid curves present ML-MCDTHX data. In ML-MCDTHX, we used $D=5, 7, 6$ orbitals for $N=15, 26, 40$, respectively. Dots, squares, triangles showcase IM-SRG results. Dashed lines are added to guide the eye. In panel d) we show phase fluctuations for $N=26$ for different numbers of orbitals in ML-MCDTHX together with the IM-SRG result. Note that already for three orbitals, the values of the energy and the density are converged for this $N$.}
\label{fig:App:Comparison}
\end{figure}

\subsection{IM-SRG vs ML-MCTDHX}

We show in Fig.~\ref{fig:App:Comparison} the energy, density and phase fluctuations of the Bose gas for different particle numbers $N$. The used values of $N$ correspond to bound, critical, and scattering states discussed in the main text. Note, however, that the box trap modifies all properties of the system if $L$ is of the order of $1/M|c|$. For example, we noticed that we need to use stronger impurity-boson interactions (and therefore larger numbers of particles) than in the main text to be able to observe significant beyond-mean-field effects. 

The ML-MCTDHX and IM-SRG results for the energy (panel a)) and the density (panel b)) are in agreement. However, phase fluctuations (panel c)) show some deviations for larger particle numbers. We notice, that while the density and the energy are accurate already for a small number of orbitals in ML-MCDTHX, phase fluctuations require more involved simulations. 
This is expected for several reasons. In particular, phase fluctuations require to determine the off-diagonal of the reduced density matrix which is a higher order observable. Note that ML-MCTDHX contains in general more information about the Hilbert space of the system in comparison to IM-SRG. Furthermore, ML-MCTDHX provides a direct access to spatially resolved observables and multicomponent settings.
In that light, ML-MCTDHX calculations of certain observables are computationally more demanding than those with IM-SRG. 

Nevertheless, increasing the number of orbitals leads to an agreement between the IM-SRG and the ML-MCDTHX results also for the phase fluctuations. 
We showcase this statement in panel d), presenting the phase fluctuations within ML-MCDTHX for an increasing orbital number in the case of $N=26$ (a similar pattern is expected for $N=40$). We observe a systematic convergence behavior.  
The main disagreement is near the boundaries of the box trap where the calculation of phase fluctuations becomes hard due to almost zero densities, see Eq.~\eqref{eq:phase_def} especially so for ML-MCTDHX which operates in first quantization. We conclude that the decrease of phase fluctuations near the boundary is a numerical artifact  caused in part by the presence of hard walls. Thus, we only show results for $x<0.35L$. 
Overall, both numerical methods predict the same behavior for the observables of interest.

\begin{figure}[t]
\centering
    \begin{subfigure}{0.5\textwidth}
    \centering
    \includegraphics[width=1\linewidth]{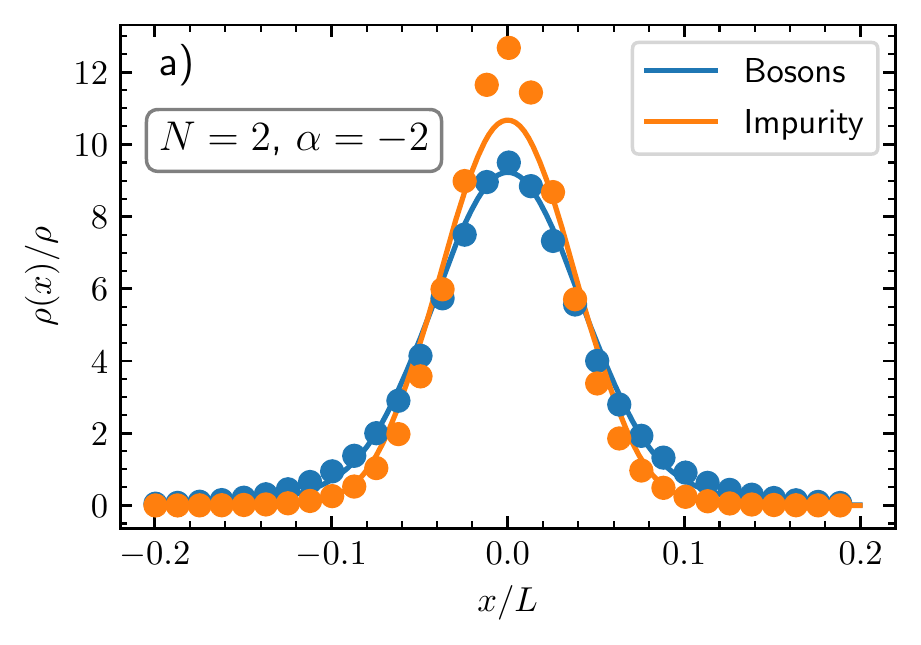}
    \end{subfigure}%
    \begin{subfigure}{0.5\textwidth}
    \centering
    \includegraphics[width=1\linewidth]{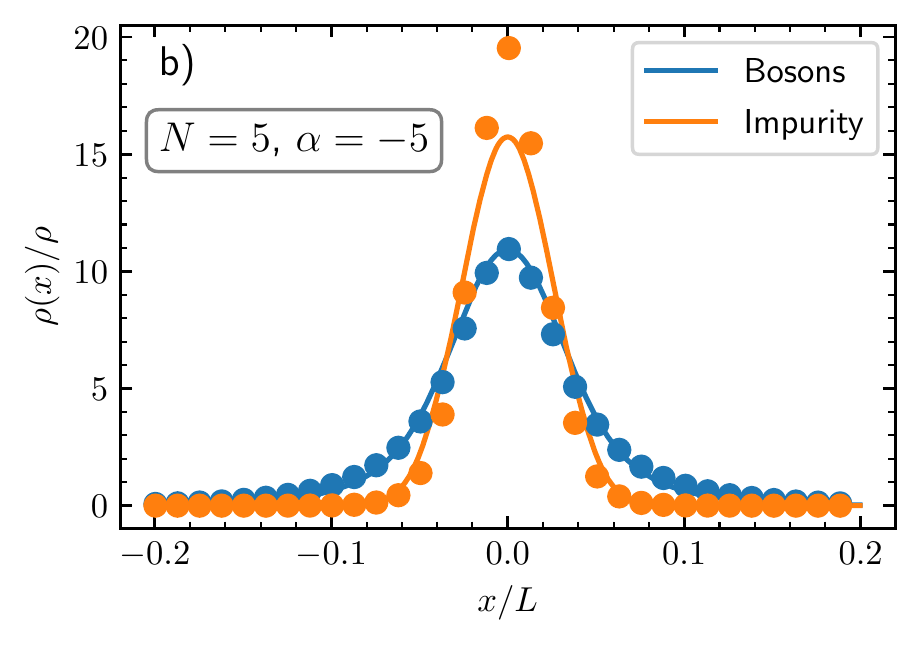}
    \end{subfigure}
    \begin{subfigure}{0.5\textwidth}
    \centering
    \includegraphics[width=1\linewidth]{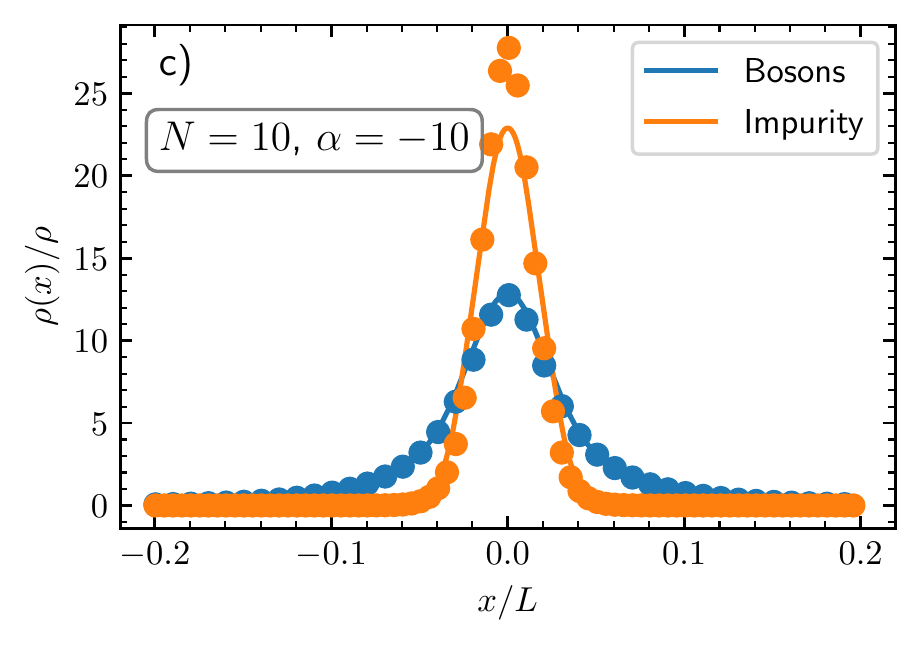}
    \end{subfigure}%
    \begin{subfigure}{0.5\textwidth}
    \centering
    \includegraphics[width=1\linewidth]{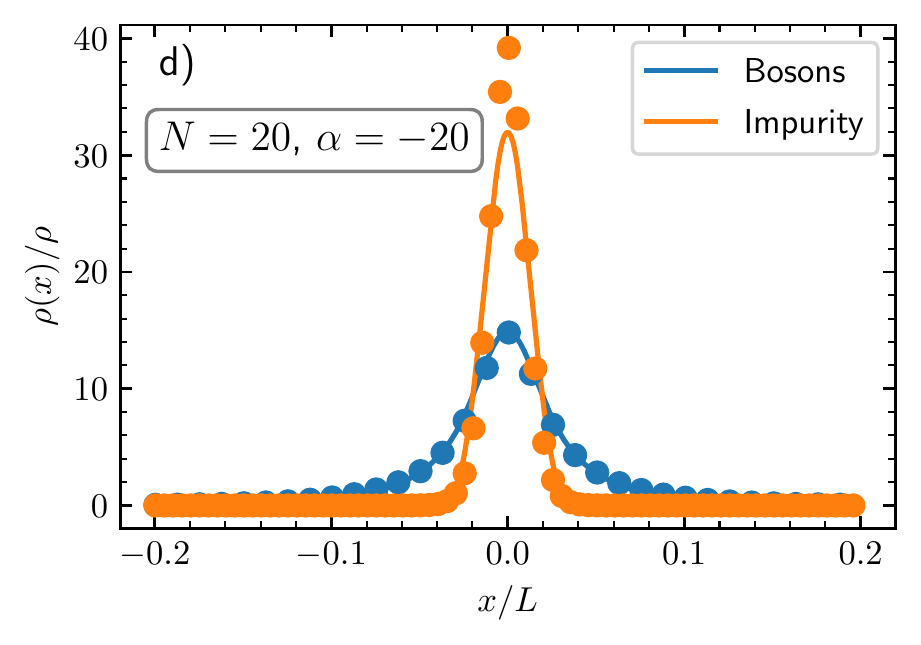}
    \end{subfigure}
    \begin{subfigure}{0.5\textwidth}
    \centering
    \includegraphics[width=1\linewidth]{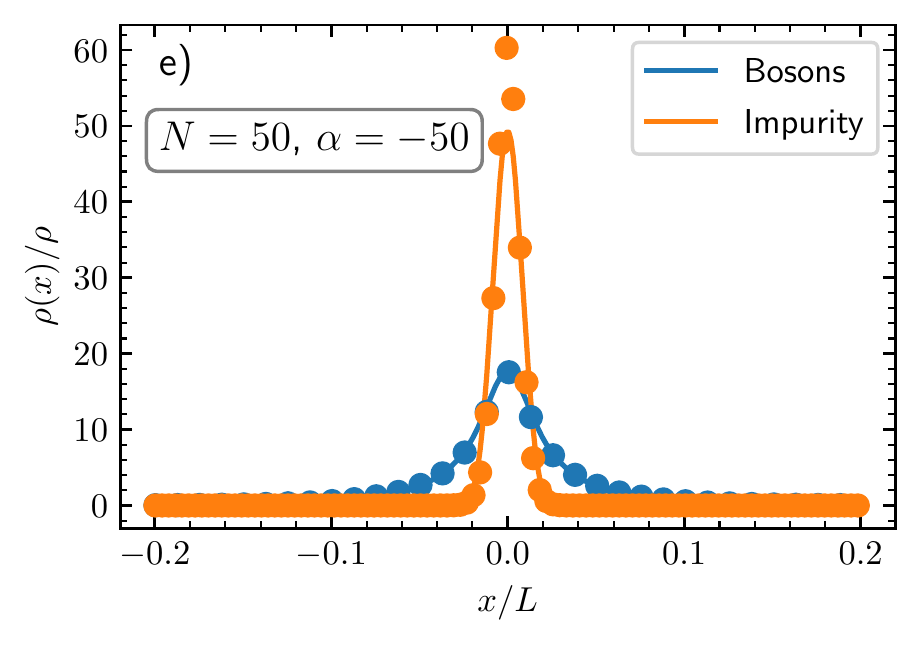}
    \end{subfigure}%
    \begin{subfigure}{0.5\textwidth}
    \centering
    \includegraphics[width=1\linewidth]{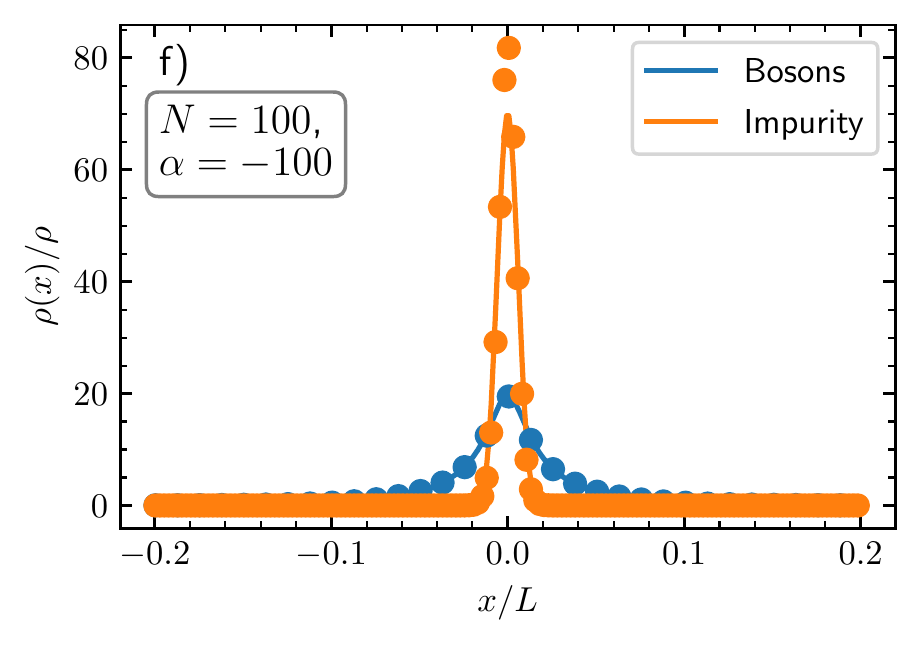}
    \end{subfigure}
\caption{Densities of the bosonic (blue) and of the impurity (orange) clouds for different particle numbers. The mass of the impurity is identical to that of a boson, i.e., $m=M$. The other parameters are chosen such that $N/\alpha=-1$. The size of the system is $LM|c|=40$. The curves (dots) are the ML-MCDTHX (mean-field) results.}
\label{fig:App:Density}
\end{figure}

\subsection{Mobile impurity: Two-component mean-field approximation}

For the sake of completeness, here, we apply the mean-field approach to the problem of a mobile impurity without the transformation to relative coordinates, see Eq.~\eqref{eq: relCoord}. We numerically solve the following set of coupled Gross-Pitaevskii equations 
\begin{equation}
\begin{split}
    &\bigg[ -\frac{\hbar}{2M} \frac{d^2}{dx^2} + g (N-1) |\Psi_B(x)|^2+c |\Psi_I(y)|^2 \bigg] \Psi_B(x)=\mu_B\Psi_B(x), \\
    &\bigg[ -\frac{\hbar}{2m} \frac{d^2}{dy^2} + c |\Psi_B(x)|^2 \bigg] \Psi_I(y)=\mu_I\Psi_I(y),
\end{split}
\end{equation}
for increasing particle numbers while fixing the ratio $N/\alpha=-1$. 
Here, $\Psi_I$ ($\Psi_B$) is the mean-field wave function of the impurity (bosons). Below, we assume $m/M=1$.

To justify this mean-field approximation, we benchmark it against ML-MCDTHX. [Note that we cannot use  the current implementation of IM-SRG for such a benchmark, as it cannot be used to study multicomponent systems.] Our findings are illustrated in Fig.~\ref{fig:App:Density} where the one-body densities of the impurity and the bosons are shown.
It becomes evident that a larger particle number results in a higher bosonic density at the position of the impurity as the effective boson-boson interaction decreases if the ratio $N/\alpha$ is kept fixed.

The considered weak boson-boson interactions lead to a good agreement between the mean-field and ML-MCTDHX methods, at least for the density of the Bose gas. This is expected since for these weak interactions boson-boson correlations are suppressed. 
For the impurity, the deviations between the ML-MCDTHX and the mean-field predictions become more noticeable for the largest numbers of bosons. Particularly, the impurity appears to be more spatially localized in the mean-field approach. 

Our data allow us to conclude that the mean-field approximation is able to provide adequate results also without transformation to a co-moving frame. However, such a transformation is needed to obtain some analytic insight into the system, as we discuss in the main text. If one is simply interested in estimating lower order observables such as densities in the mean-field approximation, then it seems that it is sufficient to work in the laboratory frame.

\section{Mean-field Solution for Hard-Wall Boundary Conditions}
\label{App:Closed}

To complement the mean-field studies in the main text, here, we  present a solution of Eq.~\eqref{eq:GPE} for a box trap, i.e., for $f(-L/2)=f(L/2)=0$. For these boundary conditions, the solution that becomes the many-body bound state in the limit $L\to\infty$ is given by the Jacobi-cs function~\cite{abramowitz1972handbook}:
\begin{equation}
\label{eq:MFclosed}
    f(x)=\sqrt{\frac{4K(p)^2}{M g L^2\delta^2(N-1)}}\,\text{cs}\left( \frac{2K(p)}{\delta}\left[\frac{|x|}{L}+\frac{\delta-1}{2}\right], p \right),
\end{equation}
with
\begin{equation}
\mu=2 K(p)^2\frac{p-2}{M\delta^2L^2}.
\end{equation}

To find the parameters $p$  and $\delta$, one should use the normalization condition and the boundary condition due to the delta-function potential:
\begin{align}
     \int |f|^2(x)\mathrm{d}x=1, \qquad \frac{\partial f}{\partial x}\bigg|_{x=0^+}=-cMf(0).
\end{align}

Note that there are other Jacobi-elliptic functions that can solve the GPE, for example, the Jacobi-sc function:
\begin{equation}
    f(x)=\sqrt{\frac{4K(p)^2(1-p)}{M gL^2\delta^2(N-1)}}\,\text{sc}\left( 2K(p)\left[ -\frac{|x|}{\delta L} + \frac{1}{2\delta}\right], p \right),
\end{equation}
with
\begin{equation}
\mu=2 K(p)^2\frac{p-2}{\delta^2L^2}.
\end{equation}
This function, however, does not lead to a physical solution in the limit $L\to\infty$, and therefore we do not consider it here. We refrain from discussing any further solutions, which may, for example, correspond to the scattering solution, Eq.~\eqref{eq:MFscatt}, from the main text. It turns out that hard-wall boundary conditions make it harder to find correct solutions for systems with finite $L$. 

\section{Zero-Density Limit within Mean-field Approximation}
\label{App:DetailsZeroDensity}

In this Appendix, we provide some technical details for the results presented in Sec.~\ref{chap:Expand}. 

\subsection{Many-body bound state}
We first of all notice that the solution from Eq.~\eqref{eq:MFmb} presented in the main text for periodic boundary conditions and Eq.~\eqref{eq:MFclosed} from the previous appendix for closed boundary conditions are identical in the limit of $L\to\infty, p\to1$:
\begin{equation}
    f(x)=\sqrt{\frac{1}{M g(N-1)}}\frac{\ln(a)}{\delta L}\sinh\left(\ln(a)\left[\frac{|x|}{\delta L}+b\right]\right)^{-1},
\end{equation}
with $a:=\dfrac{16}{1-p}$ and $b:=\dfrac{\delta-1}{2\delta}$. We used that $K(p)\to\ 1/2\ln(a)$ \cite{abramowitz1972handbook}.

Now we need to fulfill the boundary condition due to the delta-function potential
\begin{align}
     -f'(0^+)&=Mc f(0)\\
    \Rightarrow \frac{x+1}{x-1}&=\coth(\ln(a)b)=\frac{-cM\delta L}{\ln(a)},\quad\text{ for }x:=a^{2b}
\end{align}
and normalization
\begin{align}
    1&=\lim_{L\to\infty}\int\limits_{-L/2}^{L/2}f(x)^2\mathrm{d}x\\
    \Rightarrow 1&=\frac{4}{M g(N-1)}\frac{\ln(a)}{\delta L}\left[\dfrac{1}{1-a}-\dfrac{1}{1-x}\right]\stackrel{a\gg1}{=}\frac{4}{\kappa g(N-1)}\dfrac{\ln(a)}{\delta L}\dfrac{1}{x-1}.
\end{align}

Combining these two equations leads to
\begin{equation}
    \frac{\ln(a)}{\delta L}=\frac{Mg(N-1)}{2}\zeta\,,
    \label{eq:App:BounConRes}
\end{equation}
with 
\begin{equation}
    \zeta=-\frac{2c}{g(N-1)-1}=\frac{N_\mathrm{cr}-N}{N-1}.
\end{equation}
Note that by definition $x>1$, therefore, we derive the condition for the existence of the solution:
\begin{equation}
    N\leq2\frac{|c|}{g}+1,
\end{equation}
which is in agreement with the PoT condition Eq.~\eqref{eq: DropCon} from the main text.

For the chemical potential, we derive:
\begin{equation}
    \mu=-\frac{\ln(a)^2}{2M\delta^2L^2}=-\frac{M g^2(N-1)^2}{8}\zeta^2=-\frac{1}{2Mx_\mathrm{mbb}^2}\,,
\end{equation}
with $x_\mathrm{mbb}=\frac{1}{M|c|}\frac{\zeta+1}{\zeta}$, the characteristic width defined in the main text.
For the energy per particle, we find
\begin{equation}
\begin{split}
    E/N=\lim_{L\to\infty}\mu-\frac{g(N-1)}{2}\int\limits_{-L/2}^{L/2}\mathrm{dx}f(x)^4&\stackrel{a\gg1}{=}-Mg^2(N-1)^2\left(\frac{\zeta(\zeta+1)}{8}+\frac{1}{24}\right)\\
    &=-\frac{1}{Mx_\mathrm{mbb}^2\zeta^2}\left(\frac{\zeta(\zeta+1)}{2}+\frac{1}{6}\right).
\end{split}
\end{equation}

For the function $f(x)$, we can use Eq.~\eqref{eq:App:BounConRes} to simplify the solution as
\begin{equation}
\begin{split}
    f(x)&=\sqrt{\frac{Mg(N-1)}{4}}\zeta\sinh\left( \frac{Mg(N-1)}{2}\zeta x + \mathrm{ln}(\sqrt{2\zeta+1}) \right)^{-1}\\
    &=\sqrt{\frac{2\zeta(2\zeta+1)}{x_{\mathrm{mbb}}}}\frac{1}{(2\zeta+1)e^{x/x_{\mathrm{mbb}}}-e^{-x/x_{\mathrm{mbb}}}},
\end{split}
\end{equation}
At the point of transition, $\zeta\to0$, this function can be expanded around $x\zeta\to0$ as follows
\begin{equation}
    f(x)=\sqrt{\frac{|c|M }{2}}\frac{1}{|c|Mx+1}\,.
\end{equation}

\subsection{Point of transition}
The solution with the critical particle number still supporting a many-body bound state reads
\begin{equation}
    f(x)=\sqrt{\frac{\pi^2}{M gL^2\delta^2(N-1)}}\frac{1}{\cos\left(\frac{\pi (x -L/2)}{\delta L}\right)}\,.
\end{equation}
In the limit $L\to\infty$, we have $\frac{\pi x}{\delta L}\to0$ and $\delta\to1$. Therefore, we can expand the solution such that:
\begin{equation}
    f(x)\approx\sqrt{\frac{|c|M}{2}}\frac{1}{|x|+\frac{L\delta(\delta-1)}{2}}.
\end{equation}
From the normalization condition,
\begin{equation}
    \frac{Mg\delta L(N-1)}{2\pi}=\tan\left(\frac{\pi}{2\delta}\right),
\end{equation}
we derive
\begin{equation}
    \tan\left(\frac{\pi}{2\delta}\right)\stackrel{\delta\to1}{\approx}\frac{2}{\delta-1},
\end{equation}
which leads to
\begin{equation}
    f(x)=\sqrt{\frac{|c|M }{2}}\frac{1}{|c|Mx+1}\,.
\end{equation}
Note, that this expression coincides with the one derived in the previous subsection. The corresponding chemical potential vanishes:
\begin{equation}
    \mu=\lim_{L\to\infty}\frac{\pi^2}{2M\delta^2L^2}=0.
\end{equation}

\section{Few-body Limit of the Artificial Atom}
\label{App:Few-body_Limit}

\begin{figure}[t]
\centering
    \begin{subfigure}{0.5\textwidth}
    \centering
    \includegraphics[width=1\linewidth]{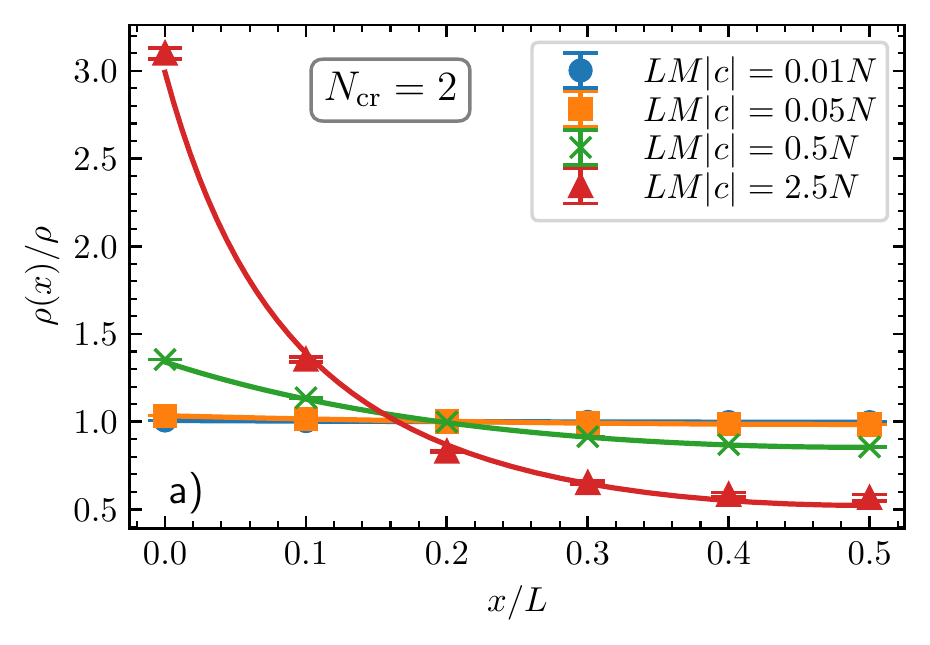}
    \end{subfigure}%
    \begin{subfigure}{0.5\textwidth}
    \centering
    \includegraphics[width=1\linewidth]{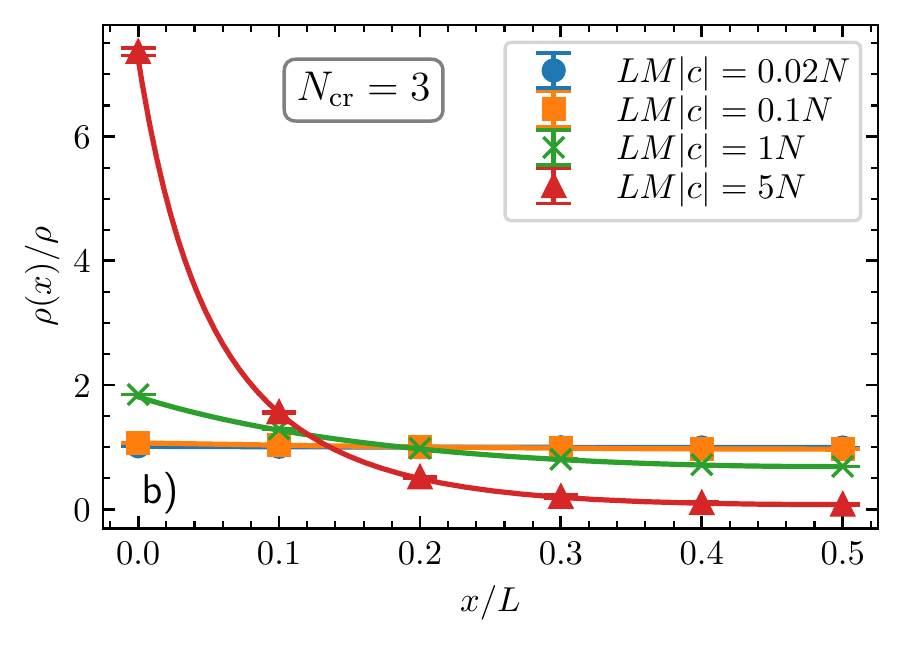}
    \end{subfigure}
    \begin{subfigure}{0.5\textwidth}
    \centering
    \includegraphics[width=1\linewidth]{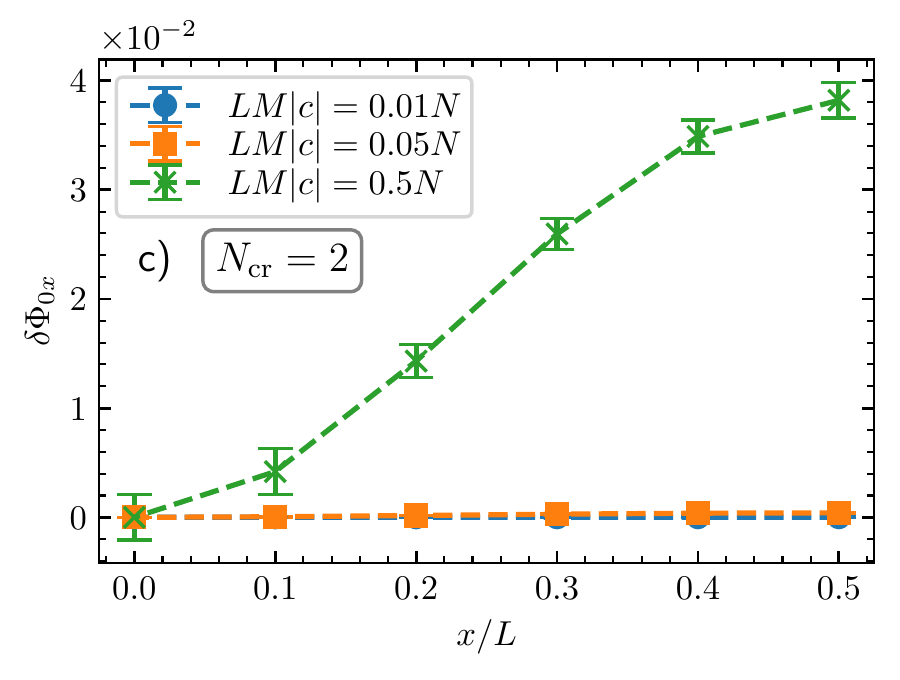}
    \end{subfigure}%
    \begin{subfigure}{0.5\textwidth}
    \centering
    \includegraphics[width=1\linewidth]{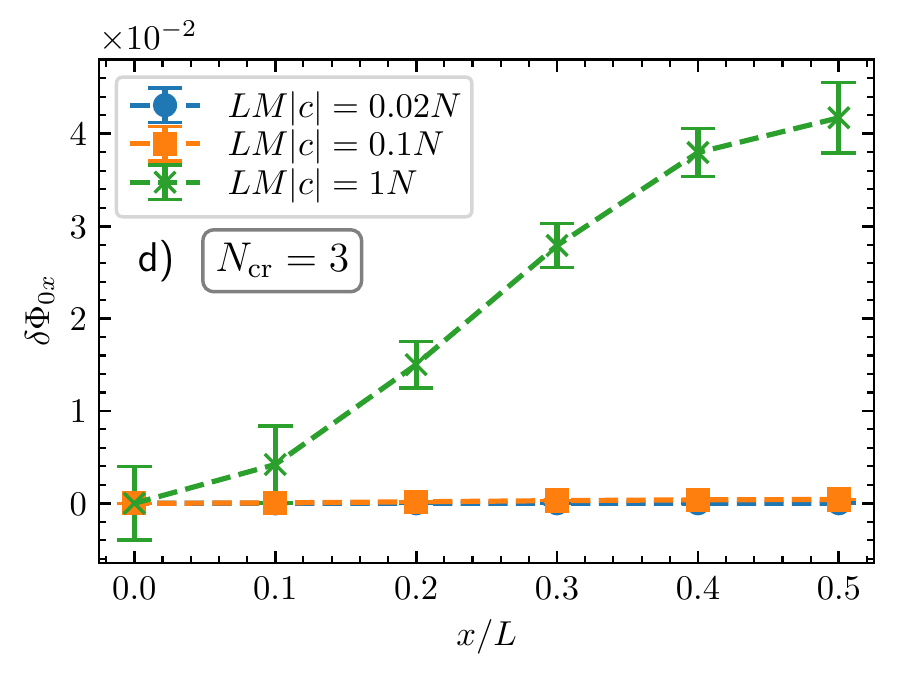}
    \end{subfigure}
\caption{Panels a) and b) show the density of the Bose gas. Circles, boxes, crosses and triangles are calculated with IM-SRG, and the solid curves are the corresponding mean-field results (Eqs.~\eqref{eq:MFmb},~\eqref{eq: MFp0}). Panels c) and~d) demonstrate phase fluctuations whose non-zero values reveal the presence of beyond-mean-field correlations. The dashed curves are provided to guide the eye. The data show results for two bosons ($N=2$) for different ring sizes. Panels a) and c) are for systems with $\alpha=-0.5$ ($N_\mathrm{cr}=2$), and panels b) and d) for $\alpha=-1$ ($N_\mathrm{cr}=3$). The numerical error bars are calculated according to the prescription given in \ref{App:NumMethods}.}
\label{fig:FewBody:densityphase}
\end{figure}

In this appendix, we discuss the smallest non-trivial system with $N=2$ assuming that only two or three bosons can be bound to the impurity. Note that {\it a priori} it is  not clear that the mean-field solution is applicable to such a few-body system.

For a two-particle system, the IM-SRG becomes essentially exact (there is only an error due to the finite Hilbert space, see \ref{App:NumMethods}, which can be easily controlled). This allows us to benchmark our mean-field results for large ring sizes. The results are presented in Fig.~\ref{fig:FewBody:densityphase} where panels~a) and~b) show the density of the Bose gas for the cases where $N_\mathrm{cr}=2$ and $N_\mathrm{cr}=3$.  It can be readily seen that independently of the ring size the behavior of the density is in an excellent agreement between the mean-field and IM-SRG approach. The same holds for the corresponding phase fluctuations depicted in panels~c) and~d) for distinct sizes of the ring. Note that the density for the largest ring size, i.e. $LM|c|=2.5N$, in both panels is too low to render meaningful values of phase fluctuations (cf. Eq.~\eqref{eq:phase_def}). However, even for the largest ring size, phase fluctuations are still low and the Bose gas can be adequately approximated with the mean-field ansatz. This numerical observation shows that the physical picture given in Sec. \ref{subsec:phys_int} is accurate even for the smallest set-ups. 

\clearpage

\printbibliography

@article{Naidon:2016dpf,
    author = "Naidon, Pascal and Endo, Shimpei",
    title = "{Efimov Physics: a review}",
    eprint = "1610.09805",
    archivePrefix = "arXiv",
    primaryClass = "quant-ph",
    doi = "10.1088/1361-6633/aa50e8",
    journal = "Rept. Prog. Phys.",
    volume = "80",
    number = "5",
    pages = "056001",
    year = "2017"
}

@article{Braaten:2004rn,
    author = "Braaten, Eric and Hammer, H.~W.",
    title = "{Universality in few-body systems with large scattering length}",
    eprint = "cond-mat/0410417",
    archivePrefix = "arXiv",
    reportNumber = "INT-PUB-04-27",
    doi = "10.1016/j.physrep.2006.03.001",
    journal = "Phys. Rept.",
    volume = "428",
    pages = "259--390",
    year = "2006"
}

@article{Efimov:1970zz,
    author = "Efimov, V.",
    title = "{Energy levels arising form the resonant two-body forces in a three-body system}",
    doi = "10.1016/0370-2693(70)90349-7",
    journal = "Phys. Lett. B",
    volume = "33",
    pages = "563--564",
    year = "1970"
}

@article{Gunn,
	doi = {10.1088/0143-0807/9/1/009},
	url = {https://doi.org/10.1088%2F0143-0807%2F9%2F1%2F009},
	year = 1988,
	% month ={01},
	publisher = {{IOP} Publishing},
	volume = {9},
	number = {1},
	pages = {51--54},
	author = {J C Gunn and J M F Gunn},
	title = {An exactly soluble Hartree problem in an external potential},
	journal = {European Journal of Physics}
}

@book{abramowitz1972handbook,
  address = {New York},
  edition = {9. Dover print., (correction reprint)},
  editor = {Abramowitz, Milton and Stegun, Irene A.},
  hdsurl = {https://hds.hebis.de/ulbda/Record/HEB124339573},
  isbn = {0486612724},
  title = {Handbook of mathematical functions : with formulas, graphs, and mathematical tables},
  uniqueid = {HEB124339573},
  url = {http://scans.hebis.de/HEBCGI/show.pl?12433957_toc.pdf},
  year = {1972}
}

@article{Hergert2016,
title = "The In-Medium Similarity Renormalization Group: A novel ab initio method for nuclei",
journal = "Physics Reports",
volume = "621",
pages = "165 - 222",
year = "2016",
note = "Memorial Volume in Honor of Gerald E. Brown",
issn = "0370-1573",
doi = "https://doi.org/10.1016/j.physrep.2015.12.007",
url = "http://www.sciencedirect.com/science/article/pii/S0370157315005414",
author = "H. Hergert and S.K. Bogner and T.D. Morris and A. Schwenk and K. Tsukiyama",
}

@book{Kehrein2006,
publisher={Springer},
  address = {Berlin},
  author = {Kehrein, Stefan},
  hdsurl = {https://hds.hebis.de/ulbda/Record/HEB192073052},
  isbn = {9783540340683},
  title = {The Flow Equation Approach to Many-Particle Systems},
  uniqueid = {HEB192073052},
  url = {https://doi.org/10.1007/3-540-34068-8},
  volume = 217,
  year = 2006
}

@article{Tsukiyama2011,
  title = {In-Medium Similarity Renormalization Group For Nuclei},
  author = {Tsukiyama, K. and Bogner, S. K. and Schwenk, A.},
  journal = {Phys. Rev. Lett.},
  volume = {106},
  issue = {22},
  pages = {222502},
  numpages = {4},
  year = {2011},
  month = {06},
  publisher = {American Physical Society},
  doi = {10.1103/PhysRevLett.106.222502},
  url = {https://link.aps.org/doi/10.1103/PhysRevLett.106.222502}
}

@article{Kane1967,
  title = {Long-Range Order in Superfluid Helium},
  author = {Kane, J. W. and Kadanoff, L. P.},
  journal = {Phys. Rev.},
  volume = {155},
  issue = {1},
  pages = {80--83},
  numpages = {0},
  year = {1967},
  publisher = {American Physical Society},
  doi = {10.1103/PhysRev.155.80},
  url = {https://link.aps.org/doi/10.1103/PhysRev.155.80}
}

@article{NoBECHohenberg,
  title = {Existence of Long-Range Order in One and Two Dimensions},
  author = {Hohenberg, P. C.},
  journal = {Phys. Rev.},
  volume = {158},
  issue = {2},
  pages = {383--386},
  numpages = {0},
  year = {1967},
  month = {06},
  publisher = {American Physical Society},
  doi = {10.1103/PhysRev.158.383},
  url = {https://link.aps.org/doi/10.1103/PhysRev.158.383}
}

@article{Popov1972,
author={Popov, V.N.},
title={On the theory of the superfluidity of two- and one-dimensional bose systems},
journal={Theor Math Phys},
volume={11},
pages={565–573},
year={1972},
url={https://doi.org/10.1007/BF01028373}
}

@article{Mukherjee2017,
  title = {Homogeneous Atomic Fermi Gases},
  author = {Mukherjee, Biswaroop and Yan, Zhenjie and Patel, Parth B. and Hadzibabic, Zoran and Yefsah, Tarik and Struck, Julian and Zwierlein, Martin W.},
  journal = {Phys. Rev. Lett.},
  volume = {118},
  issue = {12},
  pages = {123401},
  numpages = {5},
  year = {2017},
  month = {03},
  publisher = {American Physical Society},
  doi = {10.1103/PhysRevLett.118.123401},
  url = {https://link.aps.org/doi/10.1103/PhysRevLett.118.123401}
}

@article{Bell_2016,
	doi = {10.1088/1367-2630/18/3/035003},
	url = {https://doi.org/10.1088/1367-2630/18/3/035003},
	year = 2016,
	month = {03},
	publisher = {{IOP} Publishing},
	volume = {18},
	number = {3},
	pages = {035003},
	author = {Thomas A Bell and Jake A P Glidden and Leif Humbert and Michael W J Bromley and Simon A Haine and Matthew J Davis and Tyler W Neely and Mark A Baker and Halina Rubinsztein-Dunlop},
	title = {Bose{\textendash}Einstein condensation in large time-averaged optical ring potentials},
	journal = {New Journal of Physics}
}

@article{Hu2016,
  title = {Bose Polarons in the Strongly Interacting Regime},
  author = {Hu, Ming-Guang and Van de Graaff, Michael J. and Kedar, Dhruv and Corson, John P. and Cornell, Eric A. and Jin, Deborah S.},
  journal = {Phys. Rev. Lett.},
  volume = {117},
  issue = {5},
  pages = {055301},
  numpages = {6},
  year = {2016},
  month = {7},
  publisher = {American Physical Society},
  doi = {10.1103/PhysRevLett.117.055301},
  url = {https://link.aps.org/doi/10.1103/PhysRevLett.117.055301}
}

@article{Nielsen2001,
title = {The three-body problem with short-range interactions},
journal = {Physics Reports},
volume = {347},
number = {5},
pages = {373-459},
year = {2001},
issn = {0370-1573},
doi = {https://doi.org/10.1016/S0370-1573(00)00107-1},
url = {https://www.sciencedirect.com/science/article/pii/S0370157300001071},
author = {E. Nielsen and D.V. Fedorov and A.S. Jensen and E. Garrido}
}

@ARTICLE{Mistakidis2022,
       author = {{Mistakidis}, S.~I. and {Volosniev}, A.~G. and {Barfknecht}, R.~E. and {Fogarty}, T. and {Busch}, Th. and {Foerster}, A. and {Schmelcher}, P. and {Zinner}, N.~T.},
        title = "{Cold atoms in low dimensions -- a laboratory for quantum dynamics}",
         year = 2022,
        month = feb,
          eid = {arXiv:2202.11071},
}

@article{Jorgensen2016,
  title = {Observation of Attractive and Repulsive Polarons in a Bose-Einstein Condensate},
  author = {J\o{}rgensen, Nils B. and Wacker, Lars and Skalmstang, Kristoffer T. and Parish, Meera M. and Levinsen, Jesper and Christensen, Rasmus S. and Bruun, Georg M. and Arlt, Jan J.},
  journal = {Phys. Rev. Lett.},
  volume = {117},
  issue = {5},
  pages = {055302},
  numpages = {6},
  year = {2016},
  month = {7},
  publisher = {American Physical Society},
  doi = {10.1103/PhysRevLett.117.055302},
  url = {https://link.aps.org/doi/10.1103/PhysRevLett.117.055302}
}

@article{Lebrat2019,
  title = {Quantized Conductance through a Spin-Selective Atomic Point Contact},
  author = {Lebrat, Martin and H\"ausler, Samuel and Fabritius, Philipp and Husmann, Dominik and Corman, Laura and Esslinger, Tilman},
  journal = {Phys. Rev. Lett.},
  volume = {123},
  issue = {19},
  pages = {193605},
  numpages = {5},
  year = {2019},
  publisher = {American Physical Society},
  doi = {10.1103/PhysRevLett.123.193605},
  url = {https://link.aps.org/doi/10.1103/PhysRevLett.123.193605}
}

@article{Hellweg2001,
   title={Phase fluctuations in Bose–Einstein condensates},
   volume={73},
   ISSN={1432-0649},
   url={http://dx.doi.org/10.1007/s003400100747},
   DOI={10.1007/s003400100747},
   number={8},
   journal={Applied Physics B},
   publisher={Springer Science and Business Media LLC},
   author={Hellweg, D. and Dettmer, S. and Ryytty, P. and Arlt, J.J. and Ertmer, W. and Sengstock, K. and Petrov, D.S. and Shlyapnikov, G.V. and Kreutzmann, H. and Santos, L. and et al.},
   year={2001},
   month={12},
   pages={781–789}
}

@article{Petrov2000,
  title = {Regimes of Quantum Degeneracy in Trapped 1D Gases},
  author = {Petrov, D. S. and Shlyapnikov, G. V. and Walraven, J. T. M.},
  journal = {Phys. Rev. Lett.},
  volume = {85},
  issue = {18},
  pages = {3745--3749},
  numpages = {0},
  year = {2000},
  month = {10},
  publisher = {American Physical Society},
  doi = {10.1103/PhysRevLett.85.3745},
  url = {https://link.aps.org/doi/10.1103/PhysRevLett.85.3745}
}

@book{landau1977quantum,
  title={Quantum Mechanics: Non-relativistic Theory},
  author={Landau, L. D. and Lifshitz, E. M.},
  isbn={9780080190129},
  lccn={76018223},
  series={Course of theoretical physics},
  year={1977},
  publisher={Pergamon Press}
}

@article{Girardeau1960,
author = {Girardeau,M. },
title = {Relationship between Systems of Impenetrable Bosons and Fermions in One Dimension},
journal = {Journal of Mathematical Physics},
volume = {1},
number = {6},
pages = {516-523},
year = {1960},
doi = {10.1063/1.1703687},
URL = {https://doi.org/10.1063/1.1703687},
eprint = {https://doi.org/10.1063/1.1703687}
}

@article{Kolomeisky2004,
  title = {Ground-state properties of artificial bosonic atoms, Bose interaction blockade, and the single-atom pipette},
  author = {Kolomeisky, Eugene B. and Straley, Joseph P. and Kalas, Ryan M.},
  journal = {Phys. Rev. A},
  volume = {69},
  issue = {6},
  pages = {063401},
  numpages = {16},
  year = {2004},
  % month ={Jun},
  publisher = {American Physical Society},
  doi = {10.1103/PhysRevA.69.063401},
  url = {https://link.aps.org/doi/10.1103/PhysRevA.69.063401}
}

@article{Straley2007,
  title = {Interacting bosons in a nearly resonant potential well},
  author = {Straley, Joseph P. and Kolomeisky, Eugene B.},
  journal = {Phys. Rev. A},
  volume = {75},
  issue = {6},
  pages = {063421},
  numpages = {5},
  year = {2007},
  % month ={Jun},
  publisher = {American Physical Society},
  doi = {10.1103/PhysRevA.75.063421},
  url = {https://link.aps.org/doi/10.1103/PhysRevA.75.063421}
}

@article{Lieb1992,
  title = {Heavy atoms in the strong magnetic field of a neutron star},
  author = {Lieb, E. H. and Solovej, J. P. and Yngvason, J.},
  journal = {Phys. Rev. Lett.},
  volume = {69},
  issue = {5},
  pages = {749--752},
  numpages = {0},
  year = {1992},
  % month ={Aug},
  publisher = {American Physical Society},
  doi = {10.1103/PhysRevLett.69.749},
  url = {https://link.aps.org/doi/10.1103/PhysRevLett.69.749}
}

@article{Mistakidis2019,
  title = {Effective approach to impurity dynamics in one-dimensional trapped Bose gases},
  author = {Mistakidis, S. I. and Volosniev, A. G. and Zinner, N. T. and Schmelcher, P.},
  journal = {Phys. Rev. A},
  volume = {100},
  issue = {1},
  pages = {013619},
  numpages = {12},
  year = {2019},
  % month ={7},
  publisher = {American Physical Society},
  doi = {10.1103/PhysRevA.100.013619},
  url = {https://link.aps.org/doi/10.1103/PhysRevA.100.013619}
}

@article{Lode2020,
  title = {Colloquium: Multiconfigurational time-dependent Hartree approaches for indistinguishable particles},
  author = {Lode, Axel U. J. and L\'ev\^eque, Camille and Madsen, Lars Bojer and Streltsov, Alexej I. and Alon, Ofir E.},
  journal = {Rev. Mod. Phys.},
  volume = {92},
  issue = {1},
  pages = {011001},
  numpages = {21},
  year = {2020},
  publisher = {American Physical Society},
  doi = {10.1103/RevModPhys.92.011001},
  url = {https://link.aps.org/doi/10.1103/RevModPhys.92.011001}
}

@article{Jager2020,
  title = {Strong-coupling Bose polarons in one dimension: Condensate deformation and modified Bogoliubov phonons},
  author = {Jager, J. and Barnett, R. and Will, M. and Fleischhauer, M.},
  journal = {Phys. Rev. Research},
  volume = {2},
  issue = {3},
  pages = {033142},
  numpages = {8},
  year = {2020},
  publisher = {American Physical Society},
  doi = {10.1103/PhysRevResearch.2.033142},
  url = {https://link.aps.org/doi/10.1103/PhysRevResearch.2.033142}
}

@article{Lee1953,
  title = {The Motion of Slow Electrons in a Polar Crystal},
  author = {Lee, T. D. and Low, F. E. and Pines, D.},
  journal = {Phys. Rev.},
  volume = {90},
  issue = {2},
  pages = {297--302},
  numpages = {0},
  year = {1953},
  publisher = {American Physical Society},
  doi = {10.1103/PhysRev.90.297},
  url = {https://link.aps.org/doi/10.1103/PhysRev.90.297}
}

@article{Olshanii1998,
  title = {Atomic Scattering in the Presence of an External Confinement and a Gas of Impenetrable Bosons},
  author = {Olshanii, M.},
  journal = {Phys. Rev. Lett.},
  volume = {81},
  issue = {5},
  pages = {938--941},
  numpages = {0},
  year = {1998},
  publisher = {American Physical Society},
  doi = {10.1103/PhysRevLett.81.938},
  url = {https://link.aps.org/doi/10.1103/PhysRevLett.81.938}
}

@article{TrappingCollapse,
  title = {Trapping collapse: Infinite number of repulsive bosons trapped by a generic short-range potential},
  author = {Chen, Kun and Prokof'ev, Nikolay V. and Svistunov, Boris V.},
  journal = {Phys. Rev. A},
  volume = {98},
  issue = {4},
  pages = {041602},
  numpages = {5},
  year = {2018},
  % month ={Oct},
  publisher = {American Physical Society},
  doi = {10.1103/PhysRevA.98.041602},
  url = {https://link.aps.org/doi/10.1103/PhysRevA.98.041602}
}

@article{Simon1976,
author={B. Simon},
journal={Ann. Phys.},
volume={97},
pages={279},
year={1976}
}

@book{Griffiths2005,
author={Griffiths, David J.},
year={2005},
title={Introduction to Quantum Mechanics (2nd ed.)},
publisher={Prentice Hall}
}

@article{QuantumBlockade,
  title = {Quantum Behavior of a Heavy Impurity Strongly Coupled to a Bose Gas},
  author = {Levinsen, Jesper and Ardila, Luis A. Pe\~na and Yoshida, Shuhei M. and Parish, Meera M.},
  journal = {Phys. Rev. Lett.},
  volume = {127},
  issue = {3},
  pages = {033401},
  numpages = {6},
  year = {2021},
  % month ={7},
  publisher = {American Physical Society},
  doi = {10.1103/PhysRevLett.127.033401},
  url = {https://link.aps.org/doi/10.1103/PhysRevLett.127.033401}
}

@article{Grusdt2017,
	doi = {10.1088/1367-2630/aa8a2e},
	url = {https://doi.org/10.1088/1367-2630/aa8a2e},
	year = 2017,
	% month ={oct},
	publisher = {{IOP} Publishing},
	volume = {19},
	number = {10},
	pages = {103035},
	author = {Fabian Grusdt and Gregory E Astrakharchik and Eugene Demler},
	title = {Bose polarons in ultracold atoms in one dimension: beyond the Fr{\"o}hlich paradigm}
}

@article{Sowinski2019,
	doi = {10.1088/1361-6633/ab3a80},
	url = {https://doi.org/10.1088/1361-6633/ab3a80},
	year = 2019,
	month = {9},
	publisher = {{IOP} Publishing},
	volume = {82},
	number = {10},
	pages = {104401},
	author = {Tomasz Sowi{\'{n}}ski and Miguel {\'{A}}ngel Garc{\'i}a-March},
	title = {One-dimensional mixtures of several ultracold atoms: a review},
	journal = {Reports on Progress in Physics}
}

@article{Schafer2018,
  title = {Experimental realization of ultracold Yb-$^{7}\mathrm{Li}$ mixtures in mixed dimensions},
  author = {Sch\"afer, F. and Mizukami, N. and Yu, P. and Koibuchi, S. and Bouscal, A. and Takahashi, Y.},
  journal = {Phys. Rev. A},
  volume = {98},
  issue = {5},
  pages = {051602},
  numpages = {6},
  year = {2018},
  month = {11},
  publisher = {American Physical Society},
  doi = {10.1103/PhysRevA.98.051602},
  url = {https://link.aps.org/doi/10.1103/PhysRevA.98.051602}
}

@article{Chin2010,
  title = {Feshbach resonances in ultracold gases},
  author = {Chin, Cheng and Grimm, Rudolf and Julienne, Paul and Tiesinga, Eite},
  journal = {Rev. Mod. Phys.},
  volume = {82},
  issue = {2},
  pages = {1225--1286},
  numpages = {0},
  year = {2010},
  month = {4},
  publisher = {American Physical Society},
  doi = {10.1103/RevModPhys.82.1225},
  url = {https://link.aps.org/doi/10.1103/RevModPhys.82.1225}
}

@article{Catani2012,
  title = {Quantum dynamics of impurities in a one-dimensional Bose gas},
  author = {Catani, J. and Lamporesi, G. and Naik, D. and Gring, M. and Inguscio, M. and Minardi, F. and Kantian, A. and Giamarchi, T.},
  journal = {Phys. Rev. A},
  volume = {85},
  issue = {2},
  pages = {023623},
  numpages = {6},
  year = {2012},
  month = {02},
  publisher = {American Physical Society},
  doi = {10.1103/PhysRevA.85.023623},
  url = {https://link.aps.org/doi/10.1103/PhysRevA.85.023623}
}

@article{Volosniev2017,
	doi = {10.1088/1367-2630/aa9011},
	url = {https://doi.org/10.1088/1367-2630/aa9011},
	year = 2017,
	% month ={11},
	publisher = {{IOP} Publishing},
	volume = {19},
	number = {11},
	pages = {113051},
	author = {A G Volosniev and H-W Hammer},
	title = {Flow equations for cold Bose gases},
	abstract = {We derive flow equations for cold atomic gases with one macroscopically populated energy level. The generator is chosen such that the ground state decouples from all other states in the system as the renormalization group flow progresses. We propose a self-consistent truncation scheme for the flow equations at the level of three-body operators and show how they can be used to calculate the ground state energy of a general N-body system. Moreover, we provide a general method to estimate the truncation error in the calculated energies. Finally, we test our scheme by benchmarking to the exactly solvable Lieb–Liniger model and find good agreement for weak and moderate interaction strengths.}
}

@article{TwoImpIMSRG,
	title = {Impurities in a one-dimensional Bose gas: the flow equation approach},
	author = {F. Brauneis and H.-W. Hammer and  M. Lemeshko and A. G. Volosniev},
	journal = {SciPost Phys.},
	volume = {11},
	issue = {1},
	pages={8},
	year = {2021},
	publisher = {SciPost},
	doi = {10.21468/SciPostPhys.11.1.008},
	url = {https://scipost.org/10.21468/SciPostPhys.11.1.008}
}

@article{ArtemPolaron,
  title = {Analytical approach to the Bose-polaron problem in one dimension},
  author = {Volosniev, A. G. and Hammer, H.-W.},
  journal = {Phys. Rev. A},
  volume = {96},
  issue = {3},
  pages = {031601},
  numpages = {6},
  year = {2017},
  % month ={09},
  publisher = {American Physical Society},
  doi = {10.1103/PhysRevA.96.031601},
  url = {https://link.aps.org/doi/10.1103/PhysRevA.96.031601}
}

@article{QMCPolaron,
  title = {Quantum Monte Carlo study of the Bose-polaron problem in a one-dimensional gas with contact interactions},
  author = {Parisi, L. and Giorgini, S.},
  journal = {Phys. Rev. A},
  volume = {95},
  issue = {2},
  pages = {023619},
  numpages = {9},
  year = {2017},
  % month ={02},
  publisher = {American Physical Society},
  doi = {10.1103/PhysRevA.95.023619},
  url = {https://link.aps.org/doi/10.1103/PhysRevA.95.023619}
}

@book{Popov1983,
year = {1983},
author = {Popov, V. N.},
publisher = {D. Reidel Pub. Co.},
title = {Functional integrals in quantum field theory and statistical physics}
}

@article{GROSS_1962,
title = {Motion of foreign bodies in boson systems},
journal = {Annals of Physics},
volume = {19},
number = {2},
pages = {234-253},
year = {1962},
issn = {0003-4916},
doi = {https://doi.org/10.1016/0003-4916(62)90217-8},
url = {https://www.sciencedirect.com/science/article/pii/0003491662902178},
author = {E.P Gross}
}

@article{Guenther2021,
  title = {Mobile impurity in a Bose-Einstein condensate and the orthogonality catastrophe},
  author = {Guenther, Nils-Eric and Schmidt, Richard and Bruun, Georg M. and Gurarie, Victor and Massignan, Pietro},
  journal = {Phys. Rev. A},
  volume = {103},
  issue = {1},
  pages = {013317},
  numpages = {8},
  year = {2021},
  month = {1},
  publisher = {American Physical Society},
  doi = {10.1103/PhysRevA.103.013317},
  url = {https://link.aps.org/doi/10.1103/PhysRevA.103.013317}
}

@article{Enss2020,
  title = {Theory of a resonantly interacting impurity in a Bose-Einstein condensate},
  author = {Drescher, Moritz and Salmhofer, Manfred and Enss, Tilman},
  journal = {Phys. Rev. Research},
  volume = {2},
  issue = {3},
  pages = {032011},
  numpages = {6},
  year = {2020},
  month = {7},
  publisher = {American Physical Society},
  doi = {10.1103/PhysRevResearch.2.032011},
  url = {https://link.aps.org/doi/10.1103/PhysRevResearch.2.032011}
}

@article{PANOCHKO_2019,
title = {Mean-field construction for spectrum of one-dimensional Bose polaron},
journal = {Annals of Physics},
volume = {409},
pages = {167933},
year = {2019},
issn = {0003-4916},
doi = {https://doi.org/10.1016/j.aop.2019.167933},
url = {https://www.sciencedirect.com/science/article/pii/S0003491619301885},
author = {G. Panochko and V. Pastukhov}
}

@article{Widera2012,
  title = {Dynamics of Single Neutral Impurity Atoms Immersed in an Ultracold Gas},
  author = {Spethmann, Nicolas and Kindermann, Farina and John, Shincy and Weber, Claudia and Meschede, Dieter and Widera, Artur},
  journal = {Phys. Rev. Lett.},
  volume = {109},
  issue = {23},
  pages = {235301},
  numpages = {5},
  year = {2012},
  month = {12},
  publisher = {American Physical Society},
  doi = {10.1103/PhysRevLett.109.235301},
  url = {https://link.aps.org/doi/10.1103/PhysRevLett.109.235301}
}

@book{Pethick2002,
author={C. Pethick and H. Smith},
year={2002},
title={Bose–Einstein Condensation in Dilute Gases},
publisher={Cambridge University Press}
}

@article{cao2017unified,
  title={A unified ab initio approach to the correlated quantum dynamics of ultracold fermionic and bosonic mixtures},
  author={Cao, Lushuai and Bolsinger, V and Mistakidis, S I and Koutentakis, G M and Kr{\"o}nke, Sven and Schurer, J M and Schmelcher, Peter},
  journal={J. Chem. Phys.},
  volume={147},
  number={4},
  pages={044106},
  year={2017},
  publisher={AIP Publishing LLC}
}

@article{kronke2013non,
  title={Non-equilibrium quantum dynamics of ultra-cold atomic mixtures: the multi-layer multi-configuration time-dependent Hartree method for bosons},
  author={Kr{\"o}nke, Sven and Cao, Lushuai and Vendrell, Oriol and Schmelcher, Peter},
  journal={New J. Phys.},
  volume={15},
  number={6},
  pages={063018},
  year={2013},
  publisher={IOP Publishing}
}

@article{mistakidis2021radiofrequency,
  title={Radiofrequency spectroscopy of one-dimensional trapped Bose polarons: crossover from the adiabatic to the diabatic regime},
  author={Mistakidis, S I and Koutentakis, G M and Grusdt, F and Sadeghpour, H R and Schmelcher, P},
  journal={New J. Phys.},
  volume={23},
  number={4},
  pages={043051},
  year={2021},
  publisher={IOP Publishing}
}

@article{mistakidis2020pump,
  title={Pump-probe spectroscopy of Bose polarons: Dynamical formation and coherence},
  author={Mistakidis, S I and Katsimiga, G C and Koutentakis, G M and Busch, Th and Schmelcher, P},
  journal={Phys. Rev. Research},
  volume={2},
  number={3},
  pages={033380},
  year={2020},
  publisher={APS}
}

@article{mistakidis2020induced,
  title={Induced correlations between impurities in a one-dimensional quenched Bose gas},
  author={Mistakidis, S I and Volosniev, A G and Schmelcher, Peter},
  journal={Phys. Rev. Research},
  volume={2},
  number={2},
  pages={023154},
  year={2020},
  publisher={APS}
}

@article{mistakidis2020many,
  title={Many-body quantum dynamics and induced correlations of Bose polarons},
  author={Mistakidis, S I and Koutentakis, G M and Katsimiga, G C and Busch, Th and Schmelcher, Peter},
  journal={New J. Phys.},
  volume={22},
  number={4},
  pages={043007},
  year={2020},
  publisher={IOP Publishing}
}

@article{mistakidis2019quench,
  title={Quench dynamics and orthogonality catastrophe of Bose polarons},
  author={Mistakidis, S I and Katsimiga, G C and Koutentakis, G M and Busch, Th and Schmelcher, P},
  journal={Phys. Rev. Lett.},
  volume={122},
  number={18},
  pages={183001},
  year={2019},
  publisher={APS}
}

@misc{frenkel1934wave,
  title={Wave Mechanics, Clarendon},
  author={Frenkel, J},
  year={1934},
  publisher={Oxford}
}

\end{document}